\documentclass[11pt]{article}
\RequirePackage{amsthm,amsmath, bm}
\usepackage{graphicx,psfrag,epsf}
\usepackage{enumerate}
\usepackage{natbib}
\usepackage{url} 
\usepackage{subfigure}
\usepackage{amsbsy, amssymb, color,arydshln, float}

\newcommand{\pr}{\mathbb{P}}
\newcommand{\var}{\text{var}}

\newcommand{\bOmega}{\bm{\Omega}}

\def\ind{\begin{picture}(9,8)
     \put(0,0){\line(1,0){9}}
     \put(3,0){\line(0,1){8}}
     \put(6,0){\line(0,1){8}}
     \end{picture}
    }
\def\nind{\begin{picture}(9,8)
     \put(0,0){\line(1,0){9}}
     \put(3,0){\line(0,1){8}}
     \put(6,0){\line(0,1){8}}
     \put(1,0){{\it /}}
     \end{picture}
  }

\newcommand{\td}{\text{d}}
\newcommand{\ACE}{\tau}
\newcommand{\E}{\mathbb{E}}

\newcommand{\bX}{\bm{X}}

\theoremstyle{definition}
\newtheorem{assumption}{Assumption}
\newtheorem{theorem}{Theorem}
\newtheorem{lemma}{Lemma}
\newtheorem{example}{Example}

\newtheorem{corollary}{Corollary}
\newtheorem{proposition}{Proposition}
\newtheorem{remark}{Remark}
\newtheorem{result}{Result}

\newcommand*{\QEDB}{\hfill\ensuremath{\square}}

\begin{document}

\def\spacingset#1{\renewcommand{\baselinestretch}
{#1}\small\normalsize} \spacingset{1}


\spacingset{1.45}

 \title{\bf Identification of Causal Effects Within Principal Strata Using Auxiliary Variables}
 \date{} 
 \author{Zhichao Jiang
  and 
  Peng Ding 
  \footnote{
  Zhichao Jiang (Email: zhichaojiang@umass.edu) is Assistant Professor,
  Department of Biostatistics and Epidemiology, University of Massachusetts, Amherst. Peng Ding (Email: pengdingpku@berkeley.edu) is Associate Professor,
  Department of Statistics, University of California, Berkeley.
  }
  }
 \maketitle

\bigskip
\begin{abstract}
In causal inference, principal stratification is a framework for dealing with a posttreatment intermediate variable between a treatment and an outcome, in which the principal strata are defined by the joint potential values of the intermediate variable. Because the principal strata are not fully observable, the causal effects within them, also known as the principal causal effects, are not identifiable without additional assumptions. Several previous empirical studies leveraged auxiliary variables to improve the inference of principal causal effects. We establish a general theory for identification and estimation of the principal causal effects with auxiliary variables, which provides a solid foundation for statistical inference and more insights for model building in empirical research. In particular, we consider two commonly-used strategies for principal stratification problems: principal ignorability and the conditional independence between the auxiliary variable and the outcome given principal strata and covariates. For these two strategies, we give non-parametric and semi-parametric identification results without modeling assumptions on the outcome.  When the assumptions for neither strategies are plausible, we propose a large class of flexible parametric and semi-parametric models for identifying principal causal effects. Our theory not only ensures formal identification results of several models that have been used in previous empirical studies but also generalizes them to allow for different types of outcomes and intermediate variables. 
\end{abstract}

\noindent%

\newpage 
\section{Introduction}
\label{sec::intro}

Complications arise in causal inference with an intermediate variable between the treatment and the outcome. \citet{cochran1957analysis}, \citet{rosenbaum1984consquences} and \citet{frangakis2002principal} pointed out that naively conditioning on the observed intermediate variable does not yield valid causal interpretations in general. \citet{frangakis2002principal} proposed to use principal stratification, the joint potential values of the intermediate variable under both the treatment and control, to define subgroup causal effects, because it acts as a pretreatment covariate vector unaffected by the treatment. Principal stratification has a wide range of applications with meanings varying in different scientific contexts. In noncompliance problems where the treatment received might differ from the treatment assigned, principal stratification represents individual potential compliance behavior \citep{angrist1996identification}; in truncation-by-death problems where some units die before the measurement time point of their outcomes, principal stratification represents individual potential survival status \citep{rubin2006causal}; in surrogate evaluation problems, principal stratification helps to clarify criteria for good surrogate endpoints \citep{frangakis2002principal,gilbert2008evaluating}; in mediation analysis,  principal stratification with respect to the mediator represents different causal mechanisms  from the treatment to the outcome \citep{rubin2004direct,gallop2009mediation,elliott2010bayesian,mattei2011augmented}.
 \citet{vanderweele2008simple} and \citet{forastiere2018principal} linked the principal stratification approach with the direct and indirect effect approach, and \citet{jo2008causal} linked the principal stratification approach with structural equation model for mediation analysis. These problems with intermediate variables concern average causal effects within principal strata, which are also known as the principal causal effects (PCEs).

Because we cannot simultaneously observe the potential values of the intermediate variable under the treatment and control, we do not know the principal stratum of every individual, and thus cannot identify the PCEs without additional assumptions. For a binary intermediate variable, \citet{zhang2003estimation}, \citet{cheng2006bounds} and \citet{imai2008sharp} derived large sample bounds, which can be too wide to be informative. \citet{angrist1996identification}, \citet{little1998statistical}, \citet{zhang2009likelihood} and \citet{frumento2012evaluating} imposed additional structural or modeling assumptions to achieve identification. When the intermediate variable is continuous, identification becomes more difficult because of the infinitely many principal strata. To estimate the PCEs, \citet{gilbert2008evaluating} assumed parametric models and used a likelihood approach. \citet{jin2008principal}, \cite{schwartz2011bayesian}, and \citet{zigler2012bayesian} proposed different forms of parametric and semi-parametric Bayesian approaches. However, the identifiability of their models is not formally established. Without identifiability, the likelihood function may be flat over a region of some parameters, and the Bayesian inference can be sensitive to prior specifications. See \cite{gustafson2009limits} and \citet{ding2018causal} for more discussion on identifiability.

Identification is sometimes achievable with a pretreatment auxiliary variable satisfying some conditional independence assumptions. 
We focus on two categories. The first category assumes that the outcome is independent of the principal strata given the auxiliary variable. This assumption is known as {\it principal ignorability} \citep{jo2011use,ding2017principal}. Under principal ignorability, \citet{jo2009use} and \cite{stuart2015assessing} used principal scores to analyze data with one-sided noncompliance, and \citet{joffe2007defining} suggested using principal scores to estimate general causal effects within principal strata. \citet{ding2017principal} established formal identification results for PCEs with a binary intermediate variable in randomized experiments. The other category assumes the conditional independence between the outcome and the auxiliary variable within principal strata.  We will refer to this conditional independence as {\it auxiliary independence}. This assumption motivates several identification and estimation strategies in different contexts. For a binary intermediate variable, \citet{ding2011identifiability} used the baseline quality of life as an auxiliary variable for identification when evaluating the effect on the quality of life with outcomes truncated by death. Under monotonicity, \citet{mealli2013using} relaxed \citet{ding2011identifiability}'s assumptions and discussed bounds and identification of the PCEs with a binary secondary outcome. \citet{wang2017identification} extended the strategy to observational studies and relaxed monotonicity in a sensitivity analysis. In a study with multiple independent trials, \citet{jiang2016principal} used the trial number as an auxiliary variable and proposed strategies to identify the PCEs. \citet{yuan2018identifying} weakened the identification assumptions and applied the methodology to a multi-site trial in education. Similar ideas have also been used to deal with continuous intermediate variables. In assessing the effect of an HIV vaccine on infection rate through immune response, \citet{follmann2006augmented} used the baseline immune response to the rabies vaccine as an auxiliary variable.  \citet{qin2008assessing} extended this idea to deal with time-to-event endpoints under a case-cohort sampling. \citet{gilbert2008evaluating} and \citet{huang2011comparing} proposed approaches to evaluating biomarkers based on principal stratification by incorporating baseline covariates as auxiliary variables to predict the biomarkers. These strategies also provided insights for better experimental designs. In particular, \citet{gabriel2016augmented} proposed the augmented treatment run-in design and used a baseline measure as a predictor of the potential values of the intermediate variable. However, under auxiliary independence, formal identification results are established only for binary intermediate variables \citep{ding2011identifiability,mealli2013using,jiang2016principal}.

This paper discusses the identification of PCEs defined by a general intermediate variable with auxiliary variables. We first generalize the identification results under principal ignorability in \citet{ding2017principal} to general intermediate variables in both randomized experiments and observational studies, and then study the identification under auxiliary independence in various scenarios. With auxiliary independence, we establish non-parametric identification results for discrete intermediate variables and semi-parametric identification results for continuous intermediate variables. These results do not require modeling the outcome. Without principal ignorability or auxiliary independence, we propose a large class of parametric models to identify the PCEs, which has not been formally established before. Compared with models used in previous empirical studies, our models require weaker assumptions and can deal with different types of data.

Identifiability is a cornerstone for both frequentists' \citep{bickel2015mathematical} and Bayesian \citep{gustafson2015bayesian} inferences. Our results provide theoretical bases to check the identifiability of PCEs before analyzing data. Practitioners can use our results to guide model building for principal stratification problems. Our results imply that some existing models are identifiable but some are not \citep[e.g.][]{follmann2006augmented, gilbert2008evaluating,zigler2012bayesian}. Moreover, our results reveal that some existing models invoked unnecessary assumptions for identification, for example, restricting the parameter space or imposing informative priors, although these assumptions make finite-sample inference more convenient.


The paper uses the following notation. Let {\rm i.i.d.} denote ``independently and identically distributed,'' $A\ind B \mid C$ denote the conditional independence of $A$ and $B$ given $C$, and $A \overset{\text{d}}{=} B $ denote that $A$ has the same distribution as $B$.
Let $\bm{1}(\cdot)$ be the indicator function, $\pr(\cdot)$ be the probability mass or density function,  and  $\Phi(\cdot)$ be the cumulative distribution function of the standard Normal distribution. 
We say that functions $\{ f_1(x), \ldots, f_J(x) \}$ are {\it linearly independent} if $c_0 +c_1f_1(x)+\cdots+c_Jf_J(x)=0$ for all $x$ implies $c_0=c_1=\cdots=c_J=0$. We say that a family $\mathcal{Q}$ of probability distributions is {\it complete} if 
$
\int f(v) Q( \td v)=0
$
for all $Q \in \mathcal{Q} $ implies
$
f(v)=0 ,
$
a.s. \citep{lehmann2006testing}.

\section{Notation and Assumptions}\label{sec::notation}

Let $Z$ be a binary treatment indicator with $Z=1$ for the treatment and $0$ for the control, $Y$ be an outcome of interest, and $S$ be an intermediate variable between the treatment and outcome. Let $S_{iz}$ and $Y_{iz}$ be the potential values of the intermediate variable and the outcome if unit $i$ were to receive treatment $z$ ($z=0, 1$). The observed values of the intermediate variable and the outcome are $S_i  = Z_iS_{i1} + (1-Z_i)S_{i0}$ and $Y_i = Z_iY_{i1} + (1 - Z_i)Y_{i0}$. Assume that $\{Z_i,S_{i1}, S_{i0}, Y_{i1},Y_{i0}: i=1, \ldots ,n\}$ are {\rm i.i.d.} samples drawn from an infinite superpopulation, and thus the observed $\{Z_i,S_i, Y_i: i=1, \ldots ,n\}$ are also {\rm i.i.d}. As a result, we drop the subscript $i$ for notational simplicity when no confusion would arise.

\citet{frangakis2002principal} defined principal stratification as $U_i = (S_{i1}, S_{i0})$, the joint potential values of the intermediate variable, and the PCEs as
$$
\ACE_{s_1 s_0} = \E\{ Y_{1}-Y_{0}\mid U=(s_1,s_0)\}
$$
for all $s_1,s_0$. The PCEs are not identifiable because $U$ is latent in general. It is common to exploit a pretreatment auxiliary variable for identifying the PCEs. Let $W_i$ denote this variable with meanings varying in different settings. We start with the following basic assumption.

\begin{assumption}
\label{asm:randomization}
$Z \ind (Y_1,Y_0,S_1, S_0) \mid W$.
\end{assumption}

Assumption~\ref{asm:randomization} is often guaranteed by design.
In completely randomized experiments, Assumption~\ref{asm:randomization} holds because $Z \ind (Y_1,Y_0,S_1, S_0, W) $. In a multi-center experiment with $W$ being the center number, Assumption~\ref{asm:randomization} holds because $Z$ is randomized in each center.

We consider two different assumptions for identification. The first assumption is the conditional independence between the potential outcome $Y_z$ and the principal stratum $U$ given the auxiliary variable $W$.
\begin{assumption}[principal ignorability]
\label{asm:pi}
 $Y_z \ind U \mid W$ for $z=0,1$.
\end{assumption}

Assumption~\ref{asm:pi} means that given auxiliary variable $W$, the principal stratification variable is randomly assigned with respect to the potential outcomes. It requires that conditioning on the auxiliary variable there is no difference between the distributions of the potential outcomes across principal strata. Many applied researchers have invoked it to estimate the PCEs \citep{follmann2000effect,jo2009use,jo2011use, stuart2015assessing}. To make Assumptions~\ref{asm:pi} more plausible, researchers often include all pretreatment covariates in $W$.  We provide two examples below. 

\begin{example}
\citet{follmann2000effect} studied the effect of a multi-factor intervention on mortality due to coronary heart disease, where $Z$ is the indicator of the intervention and $Y$ is the survival time of the patients. One-sided noncompliance occurred in the experiment, where patients assigned to the treatment group might not actually take the treatment. Let $S$ denote the actual treatment, which can be different from $Z$. The principal stratification variable characterizes the compliance behavior of the patients. \citet{follmann2000effect} argued that potential survival time of the patients with different compliance behavior would be similar conditioning on pretreatment covariates $W$, and estimate the PCEs under principal ignorability.
\end{example}

\begin{example}
\label{ex::ding}
\citet{ding2017principal} gave an example of a randomized experiment with truncation-by-death, where $Z$ is the treatment indicator, $S$ is the binary survival status, and $Y$ is the health-related quality of life. Because the outcome is only well defined for the survived patients, the parameter of interest is the PCE within the principal stratum of the patients who would survive regardless of the treatment. 
They used all the covariates as the auxiliary variables and invoked the principal ignorability in their analysis, which means that the health-related quality of life for always survived patients would be identical to that for other patients given the covariates.
\end{example}

The second identification assumption is the conditional independence between the potential outcome $Y_z$ and the auxiliary variable $W$ given the principal stratum $U$.
\begin{assumption}[auxiliary independence]
\label{asm:exclusion}
 $Y_z \ind W \mid U$ for $z=0,1$.
\end{assumption}
 
Assumption~\ref{asm:exclusion} requires the units with different values of the auxiliary variable to be identical in the distribution of potential outcomes if they are in the same principal stratum. Under Assumption \ref{asm:randomization}, Assumption~\ref{asm:exclusion} is equivalent to $Y \ind W \mid (Z,U)$, i.e., the auxiliary variable is independent of the outcome conditional on the treatment and principal strata. Including additional pretreatment covariates can make this assumption more plausible. However, for notational simplicity, we condition on such covariates implicitly and omit them below. In many situations, Assumption~\ref{asm:exclusion} is justifiable by design. We illustrate this in the following two examples.

\begin{example}
\label{ex::follmann}
\citet{follmann2006augmented} introduced an augmented design to assess immune response in vaccine trials, where $Z$ is the indicator of an HIV vaccine injection, $S$ is the immune response to this vaccine, and $Y$ is the infection indicator. Before the randomization of $Z$, all patients receive the rabies vaccine. Let $W$ denote the immune response to the rabies vaccine, which is correlated with $S$. Because the rabies vaccine is irrelevant to the HIV infection, the potential HIV infection status should depend only on the immune response to the HIV vaccine but not the rabies vaccine. This justifies auxiliary independence, based on which \citet{follmann2006augmented} estimated the PCEs.
\end{example}

%

\begin{example}
\label{ex::jiang}
\citet{jiang2016principal} proposed approaches to identifying the PCEs by multiple independent trials, where $Z$ is the treatment indicator, $S$ is the indicator of three-year cancer reoccurrence, and $Y$ is the five-year survival status. The data are from multiple trials with the trial number denoted by $W$. \citet{jiang2016principal} argued that the principal stratification variable is a measure of physical status, and assumed that the potential survival status does not depend on the trial number $W$ given the patient's physical status. As a result, they identified the PCEs under auxiliary independence.
\end{example}

When $S$ is binary as in Example \ref{ex::jiang}, \citet{jiang2016principal} showed the identifiability of PCEs. With a general $S$ as in Example \ref{ex::follmann}, formal identification results have not been established although several parametric or semi-parametric models have been proposed in data analysis.

In the following two sections, we will give a unified theory for the identification of the PCEs with an auxiliary variable under various scenarios. The theoretical results depend on three factors: (a) whether or not the potential intermediate variable under control $S_0$ is constant, (b) whether or not the intermediate variable $S$ is discrete or continuous, and (c) whether or not Assumption~\ref{asm:pi}~or~\ref{asm:exclusion} holds. Table~\ref{tab::roadmap} presents the overview of the key results in our paper. 

\begin{table}[htp]
\caption{Roadmap of the sufficient conditions for identifying the PCEs. Note that the results with a non-constant $S_0$ require the identification of $\pr(S_1,S_0\mid W)$.}
\begin{center}
\begin{tabular}{ccccc} \hline
& Assumptions & Type of $S$ & Requirement for $W$ & Outcome model\\
 \multicolumn{5}{c}{\bf{ Constant $S_0$}} \\
Section~\ref{sec::mon:pi}&  \ref{asm:randomization},~\ref{asm:pi}~and~\ref{asm:constant} & General &No& No\\
Section~\ref{sec::const-dis-indep}&  \ref{asm:randomization},~\ref{asm:exclusion}~and~\ref{asm:constant} & Discrete&More categories than $S$ & No\\
Section~\ref{sec::const-con-indep}&  \ref{asm:randomization},~\ref{asm:exclusion}~and~\ref{asm:constant} & General& Completeness & No\\
Section~\ref{sec::const-con-noindep}&  \ref{asm:randomization}~and~\ref{asm:constant} & General& Depends on the model of $S$& Yes\\ \hline
 \multicolumn{5}{c}{ {\bf Non-constant $S_0$ }} \\
Section~\ref{sec::pi}&  \ref{asm:randomization}~and~\ref{asm:pi} & General &No& No\\
Section~\ref{sec::nonconst-dis-indep}&  \ref{asm:randomization}~and~\ref{asm:exclusion} & Discrete&More categories than $S$ & No\\
Section~\ref{sec::nonconst-con-indep}&  \ref{asm:randomization}~and~\ref{asm:exclusion} & General& Completeness & No\\
Section~\ref{sec::nonconst-con-noindep}&  \ref{asm:randomization} & General& Depends on the model of $S$& Yes\\ \hline
\end{tabular}
\end{center}
\label{tab::roadmap}
\end{table}%

\section{Constant control intermediate variable}\label{sec::constant}

In this section, we consider the cases with a constant intermediate variable under control. Under this assumption, the distribution of principal strata is identifiable, which greatly simplifies the identification strategies. We will  study the case without this assumption in the next section.

\begin{assumption}
\label{asm:constant}
$S_{i0}=c$ for all $i$, where $c$ is a constant.
\end{assumption}
In some vaccine trials \cite[e.g.,][]{follmann2006augmented, hudgens2009assessing}, Assumption \ref{asm:constant} is plausible because vaccine antigens must be present to induce a specific immune response, which is absent in the control group. For a binary $S$, Assumption \ref{asm:constant} with $c=0$ is called {\it strong monotonicity}, which holds in the one-sided noncompliance setting because individuals assigned to the control group do not have access to the treatment \citep{sommer1991estimating,imbens2015causal}.  Under Assumption \ref{asm:constant}, $S_0$ is constant, and therefore it is not necessary to include it  in $U$,  simplifying  the PCEs to
\begin{eqnarray*}
\ACE_{s_1} = \E(Y_1-Y_0\mid S_1=s_1)= \E(Y_1\mid S_1=s_1)-\E(Y_0\mid S_1=s_1).
\end{eqnarray*}
Because $S_1$ is observed in the treatment group, we can identify
\begin{eqnarray*}
\E(Y_1\mid S_1=s_1) &=& \frac{\E\{ Y_1\bm{1}(S_1=s_1)\}}{\pr(S_1=s_1)}=\frac{ \E\left[\E\{ Y\bm{1}(S=s_1)\mid Z=1,W\}\right]}{\E\left\{\pr(S=s_1\mid Z=1,W)\right\}}
\end{eqnarray*}
 under Assumption~\ref{asm:randomization}. Thus, we need only to identify $\E(Y_0\mid S_1=s_1)$. Because $S_1$ is missing in the control group, the PCEs are not identifiable without additional assumptions. Below we will discuss the identification of PCEs under Assumption~\ref{asm:pi} or \ref{asm:exclusion}.

 \subsection{Principal ignorability}
 \label{sec::mon:pi}
 \citet{ding2017principal} identify the PCEs for a binary $S$ under principal ignorability using the principal score, which is the probability of principal strata conditional on the auxiliary variable.
 We extend its definition  to a general $S$ as,
 \begin{eqnarray*}
e_{s_1,s_0}(W) = \pr(S_1=s_1, S_0=s_0\mid W).
\end{eqnarray*}
Under Assumption~\ref{asm:constant}, the principal score simplifies to $e_{s_1}(W) = \pr(S_1=s_1\mid W)$, which is identified by $e_{s_1}(W) = \pr(S=s_1\mid Z=1, W)$ under Assumption~\ref{asm:randomization}. 
The proportions of principal strata are then identified by $e_{s_1} = \pr(S_1=s_1) =\E\{e_{s_1}(W)\} $. With principal ignorability, the following theorem presents the identification results for the PCEs.

 \begin{theorem}
 \label{thm::mon:pi}
Under Assumptions~\ref{asm:randomization},~\ref{asm:pi}~and~\ref{asm:constant}, the PCEs are identified by 
 \begin{eqnarray}
 \label{eqn::pi-ace}
\ACE_{s_1} = \E \left\{ \frac{e_{s_1}(W)}{e_{s_1}} \frac{ZY}{\pi(W)}\right\}- \E \left\{ \frac{e_{s_1}(W)}{e_{s_1}} \frac{(1-Z)Y}{1-\pi(W)}\right\},
\end{eqnarray}
 where $\pi(W) = \pr(Z=1 \mid W)$ is the propensity score.
  
 \end{theorem}

 Theorem~\ref{thm::mon:pi} shows that $\E(Y_z\mid S_1=s_1)$ can be identified by the average of the outcomes in a weighted sample, with the weight depending on both the principal score and the propensity scores. The principal score accounts for the relationship between the principal stratum membership and the covariates, whereas the propensity score accounts for the relationship between the treatment and the covariates. \citet{ding2017principal}'s result holds only in randomized experiments with a binary $S$, while Theorem~\ref{thm::mon:pi} generalizes it to allow for different types of $S$ in observational studies.

Based on  Theorem~\ref{thm::mon:pi}, we can estimate the PCEs by first estimating the principal score and the propensity score and then plugging in them into~\eqref{eqn::pi-ace} with the expectation replaced by the empirical average.

\subsection{Auxiliary independence with a discrete intermediate variable}

\label{sec::const-dis-indep}
\begin{theorem}
\label{thm::mon:discrete}
Suppose $S \in \{s_1,\ldots, s_K\}$ and $W \in \{w_1,\ldots, w_L\}$. Let $M$
 denote the $K \times L$ matrix with the $(k,l)$-th element $\pr(S=s_k \mid Z=1, W=w_l)$. Under Assumptions~\ref{asm:randomization},~\ref{asm:exclusion}~and~\ref{asm:constant}, if 
$\text{rank}(M^\top M)=K$, then the PCEs are identifiable.
\end{theorem}

From Theorem \ref{thm::mon:discrete}, a necessary condition for identification is $L\geq K$, i.e., $W$ must have more categories than $S$. Because $M$ depends only on the distribution of the observed data, the condition $\text{rank}(M^\top M)=K$ is testable. The following example illustrates the identifiability for the case with binary intermediate and auxiliary variables.

\begin{example}
Consider binary $S$ and $W.$
First, from the observed distribution and Assumption~\ref{asm:randomization}, we can identify $ \theta_{sw}=\pr(S_1=s \mid W=w)=\pr(S=s \mid Z=1, W=w)$ and $\delta_w=\E(Y_0\mid W=w)=\E(Y \mid Z=0,W=w)$ for $s,w=0,1$. Second, from Assumption~\ref{asm:exclusion}, we have 
\begin{eqnarray*}
\delta_1&=& \E(Y_0 \mid S_1=1) \theta_{11}+\E(Y_0 \mid S_1=0) \theta_{01},\\
\delta_0&=& \E(Y_0 \mid S_1=1) \theta_{10}+\E(Y_0 \mid S_1=0) \theta_{00},
\end{eqnarray*}
which are two linear equations of $\E(Y_0 \mid S_1=1)$ and $\E(Y_0 \mid S_1=0)$. If the condition $\text{rank}(M^\top M)=2$ holds, or, equivalently, $S \nind W \mid Z=1$, we can uniquely solve the two linear equations and obtain 
 \begin{eqnarray*}
\E(Y_0 \mid S_1=1)&=&\frac{\delta_1 \theta_{00}- \delta_0 \theta_{01}}{\theta_{11}\theta_{00}- \theta_{10} \theta_{01}}, \quad \E(Y_0 \mid S_1=0)=\frac{\delta_1 \theta_{10}- \delta_0 \theta_{11}}{\theta_{11}\theta_{00}- \theta_{10} \theta_{01}}.
\end{eqnarray*}
Therefore, the PCEs are identifiable. 
\end{example}

\subsection{Auxiliary independence with a general intermediate variable}
\label{sec::const-con-indep}

Identification is more difficult with a continuous intermediate variable, which generates infinitely many principal strata. 
Let $\mathcal{W}$ be the support of $W$, and $\mathcal{P}_{\mathcal{W}}= \{\pr(S_1 \mid W=w): w \in \mathcal{W} \}$ be the family of probability distributions indexed by $w$.
Based on the definition of completeness, we give a sufficient condition for identification.
\begin{theorem}
\label{thm::constant:complete}
Under Assumptions~\ref{asm:randomization},~\ref{asm:exclusion}~and~\ref{asm:constant},
if $\mathcal{P}_{\mathcal{W}}$ is complete, then the PCEs are identifiable.
\end{theorem}
As discussed before, 
the key to identify the PCEs is to identify $\E(Y_0 \mid S_1)$. From Assumptions~\ref{asm:randomization}~and~\ref{asm:exclusion}, we have 
\begin{eqnarray}
 \E(Y \mid Z=0, W=w)&=&\E(Y_0 \mid W=w) \nonumber \\
 &=&\E\{\E(Y_0 \mid S_1)\mid W=w\} \nonumber \\
 &=&\int \E(Y_0\mid S_1=s)Q(\td s) \label{eqn::YZW}
\end{eqnarray}
 for any probability measure $Q(s) = \pr(S_1 \leq s \mid W=w) $ in $ \mathcal{P}_{\mathcal{W}}$. The left-hand side of~\eqref{eqn::YZW} is directly estimable from the observed data, and the distributions in $\mathcal{P}_{\mathcal{W}}$ are identified by $\pr(S_1 \mid W) = \pr(S \mid Z=1, W)$. Therefore, \eqref{eqn::YZW} is an integral equation for $\E(Y_0\mid S_1=s)$. As a result, $\E(Y_0\mid S_1=s)$ is identifiable if it can be uniquely determined by~\eqref{eqn::YZW}, which is guaranteed by the completeness of $\mathcal{P}_{\mathcal{W}}$. When $S$ is discrete, the integral in~\eqref{eqn::YZW} becomes summation, and the completeness is the same as the rank condition in Theorem~\ref{thm::mon:discrete}.

Theorem \ref{thm::constant:complete} is general but abstract. From the well-known completeness property of an exponential family \citep[][]{lehmann2006testing}, we have a more interpretable sufficient condition for identifying PCEs. 
\begin{theorem}
\label{cor::constant:exp}
Under Assumptions~\ref{asm:randomization},~\ref{asm:exclusion}~and~\ref{asm:constant}, we further assume 
\begin{eqnarray*}
\pr(S_1=s_1 \mid W=w)=h(s)g(w)\exp\{\bm{\eta}^\top(w) \bm{t}(s_1)\},
\end{eqnarray*}
where $s_1 \rightarrow \bm{t}(s_1)$ is a one-to-one mapping and $ \{ \bm{\eta}(w): w \in \mathcal{W}\}$ contains an open set in $\mathbb{R}^d$ where $d$ is the length of the vector function $\bm{\eta}(w)$. The PCEs are identifiable.
\end{theorem}
Theorem~\ref{cor::constant:exp} requires that the distribution of $S_1$ conditional on $W$ belongs to the exponential family, but it does not require any models for the potential outcome $Y_z$. Therefore, Theorem~\ref{cor::constant:exp} guarantees semi-parametric identifiability and allows for different types of outcomes. Below we give an example for Normal distributions.

\begin{corollary}
\label{prop::bivariateN}
Under Assumptions~\ref{asm:randomization},~\ref{asm:exclusion}~and~\ref{asm:constant},
if $(S_1, W)$ follows a bivariate Normal distribution, then the PCEs are identifiable. 
\end{corollary}

\begin{remark}
\label{remark::cor1}
For a binary outcome, \citet{follmann2006augmented} assumes that the outcome follows a Probit model and   $(S_1,W)$ follows a bivariate Normal distribution, which is a special case of Corollary~\ref{prop::bivariateN}. Thus, \citet{follmann2006augmented}'s model is semi-parametrically identified even without the outcome model, and his parametric outcome model is invoked only for convenience in the finite-sample inference. 
\end{remark}

To further improve the applicability of Theorem \ref{thm::constant:complete}, we review the following lemma \citep[][Lemma 4]{hu2017nonparametric} on the completeness of a class of location-scale distribution families, which works for non-exponential distributions.

\begin{lemma}
\label{lem::semipara:com}
Suppose the support of $W$ has an interior point, and $S_1 \overset{\text{d}}{=}h(W)+\sigma(W)\epsilon$ with continuously differentiable $h(w)$ and $\sigma(w)$ and $\epsilon \ind W$.  Then, 
$\mathcal{P}_{\mathcal{W}}$ is complete if the characteristic function and density function of $\epsilon$, $\phi(t)$ and  $f(\epsilon)$, satisfy
 the following conditions:
\begin{enumerate}
[(a)]
\item $0<|\phi(t)|<C \exp(-\delta|t|)$ for all $t \in \mathbb{R}$ and some constants $C,\delta>0$;
\item $f(\epsilon)$ is continuously differentiable, $\int_{-\infty}^{+\infty}|x f'(x)|\textnormal{d}x < +\infty$, and $\int_{-\infty}^{+\infty}f^2(x)\textnormal{d}x< +\infty$;
\item for any positive integer $J$, the following functions are linearly independent,
$$
\left\{f\left(\frac{x-h_1}{\sigma_1}\right),\ldots, f\left(\frac{x-h_J}{\sigma_J}\right)\right\},
$$
 where the $(h_j,\sigma_j)$'s are distinct. 
\end{enumerate}
\end{lemma}

The existence of the interior point required by Lemma \ref{lem::semipara:com} holds automatically for continuous $W$ but fails for discrete $W$. Conditions (a) and (b) in Lemma \ref{lem::semipara:com} are technical requirements on the distribution of the error term $\epsilon$. Condition (c) means that the finite location-scale mixture of the distribution of $\epsilon$ is identifiable, which holds for many distributions \citep{everitt1981finite}. For example, Appendix B.1 shows that Conditions (a)--(c) hold when $\epsilon$ follows a Normal, $t$ or Logistic distribution. Combining Theorem \ref{thm::constant:complete} with Lemma \ref{lem::semipara:com}, we obtain the following theorem for the location-scale distribution families.

\begin{theorem}
\label{thm:semipara}
Suppose that $W$ is continuous,
 Assumptions~\ref{asm:randomization},~\ref{asm:exclusion}~and~\ref{asm:constant} hold, $S_1\overset{\text{d}}{=}h(W)+\sigma(W)\epsilon$ with continuously differentiable $h(w)$ and $\sigma(w)$, and $\epsilon \ind W$. If $\epsilon$ satisfies Conditions (a)--(c) in Lemma \ref{lem::semipara:com}, then the PCEs are identifiable.
\end{theorem}

Theorem \ref{thm:semipara} guarantees the identifiability of PCEs in many models involving distributions that do not belong to an exponential family. It allows for heteroscedastic errors and enables flexible model choices. For example, if we replace the bivariate Normal distribution assumption of $(S_1, W)$ with $S_1 \mid W=w \sim N(\mu(w), \sigma^2(w))$, Theorem~\ref{cor::constant:exp} and Corollary~\ref{prop::bivariateN} cannot be applied because $\{ \bm{\eta}(w)=(1/\sigma^2(w),\mu(w)/\sigma^2(w)): w \in \mathcal{W} \}$ is a line in $\mathbb{R}^2$. However, Theorem~\ref{thm:semipara} ensures that the PCEs are still identifiable.

\subsection{Without conditional independence}
\label{sec::const-con-noindep}
The conditional independence in Assumption~\ref{asm:pi}~or~\ref{asm:exclusion}
may be violated. In Example~\ref{ex::ding}, covariates may not be sufficient to account for the difference in the health-related quality of life across principal strata, which makes Assumption~\ref{asm:pi} implausible; in Example~\ref{ex::jiang}, different centers may have different qualities of services, which makes Assumption \ref{asm:exclusion} implausible. Without conditional independence, $W$ does not help to achieve non-parametric or semi-parametric identification. One solution is to conduct sensitivity analysis, which, however, requires to use sensitivity parameters to characterize the violation of the assumptions and further requires to specify their ranges. Sensitivity analysis gives a range of estimates rather than a point estimate, and it often depends on additional model assumptions. We will not pursue this direction in this paper. 
Instead, in this subsection, we seek an alternative route to propose some parametric models for identifying the PCEs,
in which the auxiliary variable $W$ satisfies certain modeling assumptions. We can also include other covariates $\bX$ in our models, but do not require any modeling assumptions for $\bX$. So, again, we condition on $\bX$ implicitly. 
The results in this subsection ensure the identifiability of the PCEs under many models that have been used in previous empirical studies and generalize some models to account for different types of outcomes and intermediate variables.

\begin{proposition}
\label{ex::nind:linear}
Under Assumptions~\ref{asm:randomization}~and~\ref{asm:constant}, assume that
$(S_1, Y_0 ) $ follow additive models:
\begin{eqnarray}
&&S_1 =g(W)+\sigma_1(W)\epsilon_{S_1}, \label{eq::s1_prop1}\\
&&Y_0= \beta_0+\alpha S_1+\sum_{j=1}^J \beta_j f_j(W)+\sigma_2(W)\epsilon_{Y_0}, \label{eq::y0_prop1}
\end{eqnarray} 
where $\E(\epsilon_{S_1}\mid W)=\E( \epsilon_{Y_0} \mid W)=0$, and $g(w)$ and $\sigma_1(w)$ can be unknown functions. 
If $\{1, g(w), f_1(w),\ldots,f_J(w)\}$ are linearly independent, then
 the PCEs are identifiable.
\end{proposition}

We do not need to specify $g(w)$ and $\sigma_1(w)$ because they are identifiable from the observed distribution $\pr(S,W\mid Z=1)$ under Assumption~\ref{asm:randomization}. In contrast, we need to specify $f_j(w)$'s and $\sigma_2(w)$ in the model of $Y_0$.

 Intuitively, in Proposition~\ref{ex::nind:linear}, replacing $S_1$ in \eqref{eq::y0_prop1} by \eqref{eq::s1_prop1}, we obtain an additive model of $Y_0$ on $W$, and the linear independence condition allows us to disentangle the coefficients of different terms involving $W$. For example, if $g(W)$ is quadratic in $W$ in \eqref{eq::s1_prop1} and \{$J=1$, $f_1(W)=W$\} in \eqref{eq::y0_prop1}, then the linear independence assumption holds in Proposition \ref{ex::nind:linear}. However, if $g(w)$ is linear in $w$, then the linear independence assumption fails.

If $f_j(w)=0$ for all $j$, then Proposition~\ref{ex::nind:linear} becomes a special case of Theorem~\ref{thm:semipara}.
Proposition~\ref{ex::nind:linear} guarantees the identifiability of PCEs in additive models without specifying the distributions of the error terms.

In the model of $Y_0$, we require $S_1$ to have a linear form. Identification may also be possible 
for other forms of $S_1$, but will require the knowledge of the distributions of the error terms.

For binary outcomes, we show an identification result below for the Probit model.

\begin{proposition}
\label{ex::nind:probit}
 Under Assumptions \ref{asm:randomization}~and~\ref{asm:constant},
assume that $S_1$ follows an additive model with Normal error and $Y_0$ follows a Probit model:
\begin{eqnarray*}
&&S_1 =g(W)+\epsilon_{S_1},  \quad \epsilon_{S_1} \ind W, \quad \epsilon_{S_1} \sim N(0, \sigma^2),\\
&&\pr(Y_0=1\mid S_1=s,W=w)= \Phi\left\{\beta_0+\alpha s+\sum_{j=1}^J \beta_j f_j(w)\right\},
\end{eqnarray*}
where $g(w)$ can be unknown.
If $\{1,g(w), f_1(w),\ldots,f_J(w)\}$ are linearly independent, then the PCEs are identifiable.
\end{proposition}
The model of $S_1$ in Proposition~\ref{ex::nind:probit} requires the variance of the error term $\epsilon_{S_1}$ not depend on $W$, which is different from Proposition~\ref{ex::nind:linear}. Identification may also be possible with the variance depending on $W$, but will rely on the functional form of $\var(S_1\mid W)$.

\begin{remark}  
Our result is not in contrary to \citet{follmann2006augmented}.
Without Assumption~\ref{asm:exclusion}, 
\citet{follmann2006augmented} assumed a bivariate Normal distribution for $(S_1,W)$ and used the following Probit model for $Y$:
\begin{eqnarray}
\label{eqn:follmann}
\pr(Y=1\mid Z,S_1,W)= \Phi(\beta_0+\beta_1Z+\beta_2S_1+\beta_3W+\beta_4ZS_1).
\end{eqnarray}
Under Assumption \ref{asm:randomization}, \eqref{eqn:follmann} is equivalent to 
\begin{eqnarray*}
\pr(Y_1=1\mid S_1,W)&=& \Phi\{\beta_0+\beta_1+(\beta_2+\beta_4)S_1+\beta_3W\},\\
\pr(Y_0=1\mid S_1,W)&=& \Phi(\beta_0+\beta_2S_1+\beta_3W).
\end{eqnarray*}
From Proposition~\ref{ex::nind:probit}, without the model of $Y_1$, the PCEs are not identifiable because the linear independence condition is violated. The identifiability comes from the parallel model assumption that restricts the coefficients of $W$ be the same in the models of $Y_1$ and $Y_0$. 
\end{remark} 
\begin{remark}
Without the linear independence condition, researchers often use additional information on the parameters to improve identification.  Using a Bayesian approach, \citet{zigler2012bayesian} imposed informative priors on $\alpha$. In a similar setting with a time-to-event outcome, \citet{qin2008assessing} imposed the principal ignorability $Y_0 \ind S_1\mid W$, or, equivalently, $\alpha=0$.
\end{remark}

%
%

\section{Non-constant control intermediate variable}\label{sec::nonconstant}

Assumption \ref{asm:constant} does not hold in many applications. Without it, we can never simultaneously observe $S_1$ and $S_0$, and therefore it is challenging to identify the joint distribution of $(S_1,S_0)$ in the first place, let alone the PCEs. 
Below we first use a copula model for the joint distribution of $(S_1, S_0)$, and then discuss identification of the PCEs.

\subsection{A copula model for $\pr(S_1,S_0\mid W)$}
\label{sec::copula}

Under Assumption~\ref{asm:randomization}, $\pr(S_z\mid W) = \pr(S \mid Z=z,W)$, and thus we can identify the marginal distributions of $S_z$ given $W$ from the observed data. To recover the joint distribution of $(S_1,S_0)$ given $W$ from the marginal distributions, we need some prior knowledge about the association between $S_1$ and $S_0$ conditional on $W$. For a binary $S$, a commonly-used assumption to recover the joint distribution of $(S_1,S_0)$ is the monotonicity assumption, i.e., $S_1 \geq S_0$. Under this assumption, we can identify $\pr(S_1=1,S_0=1\mid W) =\pr(S=1 \mid Z=0,W) $, $\pr(S_1=0,S_0=0\mid W) =\pr(S=0 \mid Z=1,W) $ and $\pr(S_1=1,S_0=0\mid W) =\pr(S=1 \mid Z=1,W) -\pr(S=1 \mid Z=0,W) $.
For a continuous $S$, \citet{efron1991compliance} and \citet{jin2008principal} discussed the equipercentile equating assumption, i.e., $F_1(S_1 \mid W)=F_0(S_0\mid W)$, where $F_z(\cdot\mid W)$ is the cumulative distribution function of $S_z$ given $W$ for $z=0,1$. Under this assumption, $S_z$ determines $S_{1-z}$ based on $F_1(\cdot\mid W)$ and $F_0(\cdot\mid W)$ for $z=0,1$.

The monotonicity and equipercentile equating assumptions are special cases of the copula approach \citep{nelsen2007introduction}, which is a general strategy to obtain the joint distribution from marginal distributions. Various copula models have been proposed to model principal strata \citep{roy2008principal,bartolucci2011modeling,schwartz2011bayesian,daniels2012bayesian,conlon2014surrogacy, yang2018using,kim2020health}. 
 Conditional on $w$, we assume
 \begin{eqnarray}
 \label{eqn::copula}
\pr(S_1, S_0\mid W=w) = C_{\rho}\{\pr( S_1\mid W=w),\pr(S_0\mid W=w)\},
\end{eqnarray}
where $C_{\rho}(\cdot,\cdot)$ is a copula and $\rho$ is a measure of the association between $S_1$ and $S_0$.  If we know $\rho$, then we can identify $\pr( S_1, S_0\mid W=w)$ from the marginal distributions by \eqref{eqn::copula}. Otherwise, we can view $\rho$ as a sensitivity parameter and conduct sensitivity analysis by varying $\rho$. 

 \subsection{Principal ignorability}
 \label{sec::pi}
Assume that the principal score $e_{s_1,s_0}(W)=\pr(S_1,S_0 \mid W)$ is identifiable. So the density of the principal strata equals $e_{s_1,s_0}= \E \left\{ e_{s_1,s_0}(W)\right\}$.
 Similar to Section~\ref{sec::mon:pi}, we establish the identification of PCEs using the principal scores.
 \label{sec::const-pi}
 \begin{theorem}
 \label{thm::pi}
Under Assumptions~\ref{asm:randomization}~and~\ref{asm:pi}, if $e_{s_1,s_0}(W)$ is identifiable, then
 the PCEs are identified by 
 \begin{eqnarray*}
\ACE_{s_1 s_0} = \E \left\{ \frac{e_{s_1,s_0}(W)}{e_{s_1,s_0}}\cdot \frac{ZY}{\pi(W)}\right\}- \E \left\{ \frac{e_{s_1,s_0}(W)}{e_{s_1,s_0}} \cdot \frac{(1-Z)Y}{1-\pi(W)}\right\}.
\end{eqnarray*}
 \end{theorem}
 Theorem~\ref{thm::pi} generalizes Theorem~\ref{thm::mon:pi} to the case with non-constant control intermediate variables.
 It shows that $\E(Y_z\mid S_1=s_1,S_0=s_0)$ can be identified by the average of the outcomes in a weighted sample, with the weight depending on both the principal score and the propensity score. 

\subsection{Auxiliary independence with a discrete intermediate variable}

\label{sec::nonconst-dis-indep}
   
We give the identification results for discrete intermediate variables.
  
 \begin{theorem}
\label{thm::nonmon:discrete}
Suppose $S \in \{s_1,\ldots, s_K\}$, $W \in \{w_1,\ldots, w_L\}$, Assumptions \ref{asm:randomization} and \ref{asm:exclusion} hold, and $\pr(S_1,S_0 \mid W)$ is identifiable. 
\begin{enumerate}
[(a)]
\item Let $M_{s_0}$ denote the $K \times L$ matrix with $(k,l)$-th element $\pr(S_1=s_k \mid S_0=s_0,W=w_l)$. For a fixed $s_0$, if $\text{rank}(M_{s_0}^\top M_{s_0})=K$, then $\pr(Y_0 \mid S_1, S_0=s_0)$ is identifiable.
\item Let $M_{s_1}$ denote the $K \times L$ matrix with $(k,l)$-th element $\pr(S_0=s_k \mid S_1=s_1,W=w_l)$. For a fixed $s_1$, if $\text{rank}(M_{s_1}^\top M_{s_1})=K$, then $\pr(Y_1 \mid S_1=s_1, S_0)$ is identifiable.
\item If (a) and (b) above hold for all $s_1$ and $s_0$, then the PCEs are identifiable.
\end{enumerate}
\end{theorem}

 Theorem \ref{thm::nonmon:discrete} extends Theorem \ref{thm::mon:discrete}.  As special cases of Theorem \ref{thm::nonmon:discrete},  for a binary intermediate variable under monotonicity, \citet{ding2011identifiability} and \citet{jiang2016principal} gave the identification results, and the rank conditions in Theorem \ref{thm::nonmon:discrete} simplify to conditional independence relationships $S_1 \nind W \mid S_0 $ and $S_0 \nind W \mid S_1$.

\subsection{Auxiliary independence with a general intermediate variable}
\label{sec::nonconst-con-indep}

Recalling that $\mathcal{W}$ is the support of $W$. For fixed $s_0$ and $s_1$, let $\mathcal{P}_{\mathcal{W}, s_0}= \{\pr(S_1 \mid S_0=s_0, W=w): w \in \mathcal{W} \}$ and $\mathcal{P}_{\mathcal{W}, s_1}= \{\pr(S_0 \mid S_1=s_1, W=w): w \in \mathcal{W} \}$ be the families of the distributions indexed by $w$ given $s_0$ and $s_1$, respectively.

Similar to Section 3.2, the identifiability of PCEs reduces to the completeness of $\mathcal{P}_{\mathcal{W}, s_0}$ and $\mathcal{P}_{\mathcal{W}, s_1}$.

\begin{theorem}
\label{thm:noncon:complete}
Suppose that Assumptions \ref{asm:randomization} and \ref{asm:exclusion} hold, and $\pr(S_1,S_0 \mid W)$ is identifiable.
\begin{enumerate}[(a)]
\item
If $\mathcal{P}_{\mathcal{W}, s_0}$ is complete for all $s_0$, then  $\pr(Y, S_1,S_0,W\mid Z=0)$ is identifiable.
\item 
If $\mathcal{P}_{\mathcal{W}, s_1}$ is complete for all $s_1$, then $\pr(Y, S_1,S_0,W\mid Z=1)$ is identifiable.
\item If (a) and (b) above hold, then
the PCEs are identifiable.
\end{enumerate}
\end{theorem}

Similar to Theorem \ref{thm::constant:complete}, Theorem \ref{thm:noncon:complete} does not require any models for the distribution of $Y_z$ $(z=0,1)$, which guarantees  the non-parametric or semi-parametric identification of PCEs. Based on the completeness of the location-scale distribution families in Lemma \ref{lem::semipara:com}, we can obtain identification results for some widely-used models. We give an example below.

\begin{corollary}
\label{cor:trivariateN}
For a continuous $W$,
suppose that Assumptions \ref{asm:randomization} and \ref{asm:exclusion} hold. If
\begin{eqnarray}\label{eq::bivariateNormal}
(S_1,S_0)\mid W=w \sim \bm{N}_2 \left\{ 
\begin{pmatrix}
\mu_1(w)\\
\mu_0(w)
\end{pmatrix}, \begin{pmatrix}
 \sigma^2_{1}(w)& \rho(w)\sigma_{1}(w)\sigma_{0}(w)\\
\rho(w)\sigma_{1}(w)\sigma_{0}(w) & \sigma^2_0(w)
\end{pmatrix}
 \right\}
\end{eqnarray}
with a known $\rho(w)$, then
 the PCEs are identifiable. 
\end{corollary}

 Corollary \ref{cor:trivariateN} does not need any models for the outcome, but requires the auxiliary variable to be continuous.  
In Corollary~\ref{cor:trivariateN}, with a known $\rho(w)$, we can identify the joint distribution of $(S_1,S_0)$ given $W$ from the marginal distributions of $S_z$ given $W$. Therefore, the PCEs are identifiable from Theorem~\ref{thm:noncon:complete}.
To apply Corollary~\ref{cor:trivariateN}, we need to pre-specify the correlation coefficient $\rho(w)$, which is a sensitive parameter in practice.

\subsection{Without conditional independence}
\label{sec::nonconst-con-noindep}
Similar to the case with a constant control intermediate variable, 
we propose some useful parametric models for identifying the PCEs using 
the auxiliary variable $W$ when Assumptions~\ref{asm:pi}~or~\ref{asm:exclusion} fails.

\begin{proposition}
\label{ex:noncon:noind:binary}
For a binary $S$ with monotonicity $S_1 \geq S_0$, 
suppose that Assumption \ref{asm:randomization} holds, and $Y_1$ and $Y_0$ follow linear models
\begin{eqnarray}
\label{eqn:noncon:noind:binary}
\E(Y_z\mid S_1,S_0,W)=\beta_{z0}+\beta_{z1}S_1+\beta_{z2} S_0+\beta_{z3}W,\quad (z=0,1).
\end{eqnarray}
If both
\begin{eqnarray}
\label{eq::ratios}
\frac{ \pr(S=1\mid Z=1,W=w)}{\pr(S=1\mid Z=0,W=w)} \quad \text{ and } \quad
\frac{ \pr(S=0\mid Z=1,W=w)}{\pr(S=0\mid Z=0,W=w)}
\end{eqnarray}
are not constant in $w$, then the PCEs are identifiable.
\end{proposition}

We can use observed data to check whether the two terms in \eqref{eq::ratios} are constant in $w$.
For a binary $W$, the only restriction of \eqref{eqn:noncon:noind:binary} is no interaction term among $(S_1,S_0, W)$ in the model of $Y$, which is similar to some existing no-interaction or homogeneity assumption \citep{ding2011identifiability, wang2017identification}.

For a continuous intermediate variable, we give the following proposition.

\begin{proposition}
\label{ex:noncon:linear}
Suppose that Assumption \ref{asm:randomization} holds,  $(S_1,S_0 )$ given $W$ follows \eqref{eq::bivariateNormal} with a known $\rho(w)$, and $Y_1$ and $Y_0$ follow additive models:
\begin{eqnarray*}
&&Y_1= \beta_0+\alpha_1 S_1+\alpha_0 S_0+\sum_{j=1}^J \beta_j f_j(W)+\sigma^2_1(W)\epsilon_{Y_1},\\
&&Y_0= \beta'_0+\alpha_1'S_1+\alpha'_0 S_0+\sum_{j=1}^J \beta'_j h_j(W)+\sigma^{2}_0(W)\epsilon_{Y_0},\\
&&(\epsilon_{Y_1},\epsilon_{Y_2}) \ind (S_1,S_0,W).
\end{eqnarray*}
The PCEs are identifiable if the following two conditions hold:
\begin{enumerate}
[(a)]
\item
 $\{1,s_1,\E(S_0\mid S_1=s_1,W=w), f_1(w),\ldots,f_J(w)\}$ are linearly independent as functions of $(s_1,w)$;
\item
$\{1,s_0,\E(S_1\mid S_0=s_0,W=w), h_1(w),\ldots,h_J(w)\}$ are linearly independent as functions of $(s_0,w)$.
\end{enumerate}
\end{proposition}

Proposition~\ref{ex:noncon:linear}, as an extension of Proposition~\ref{ex::nind:linear}, is mostly useful for continuous outcomes. The Normality in \eqref{eq::bivariateNormal} implies a linear relation of $S_0$ on $S_1$ given $W$, i.e., $S_0= a_0(W)S_1+b_0(W)\epsilon_{S_0}$ with $a_0(w)$ and $b_0(w)$ determined by the distribution of $(S_1,S_0 )$ given $W$. Then, in Proposition~\ref{ex:noncon:linear}, we can obtain an additive model of $Y_1$ on $S_1$ and $W$ by replacing $S_0$ in the model of $Y_1$. The linear independence condition (a) allows us to disentangle the coefficients of different terms involving $S_1$ and $W$. Similar discussion applies to condition (b).

The Normality in \eqref{eq::bivariateNormal} is also helpful for binary outcomes. The following proposition gives the identification result under the Probit model for $Y_z$. 
 \begin{proposition}
 \label{ex::noncon:nind:probit}
Suppose that Assumption \ref{asm:randomization} holds, and $(S_1,S_0 )$ given $W$ follows \eqref{eq::bivariateNormal} with a known $\rho(w)$. 
Suppose $Y_1$ and $Y_0$ follow Probit models:
\begin{eqnarray} 
&&\pr(Y_1=1\mid S_1=s_1,S_0=s_0,W=w)= \Phi\left\{\beta_0+\alpha_1 s_1+\alpha_0 s_0+\sum_{j=1}^J \beta_j f_j(w)\right\}, \label{eq::y1_prop_probit} \\
&& \pr(Y_0=1\mid S_1=s_1,S_0=s_0,W=w)= \Phi\left\{\beta'_0+\alpha'_1 s_1+\alpha'_0 s_0+\sum_{j=1}^J \beta'_j h_j(w)\right\}. \label{eq::y0_prop_probit}
\end{eqnarray}  
If Conditions (a) and (b) in Proposition \ref{ex:noncon:linear} hold, then the PCEs are identifiable.
\end{proposition}

\begin{remark} 
Using a Bayesian approach, 
 \citet{zigler2012bayesian} assumed a trivariate Normal distribution for $(S_1,S_0, W)$ with a sensitivity parameter to characterize the correlation between $S_1$ and $S_0$, and Probit models for $Y_z$ with $f_j(w)$ and $h_j(w)$ linear in $w$.  Under their models, the conditional expectation $\E(S_0\mid S_1=s_1,W=w)$ is linear in both $s_1$ and $w$, and $\E(S_1\mid S_0=s_0,W=w)$ is linear in both $s_0$ and $w$. Thus, the linear independence condition is violated, and the parameters are not identifiable. To mitigate the inferential difficulties, \citet{zigler2012bayesian} imposed informative priors on $\alpha_1-\alpha'_1$ and $\alpha_0-\alpha'_0$.
 \end{remark}

\section{Numerical examples}\label{sec::examples}
 
In the frequentists' inference, non-identifiability renders the likelihood function flat over a region for some parameters, and the classical repeated sampling theory of the maximum likelihood estimates do not apply \citep{bickel2015mathematical}. Computationally, the Bayesian machinery is still applicable as long as the priors are proper. The simulation below, however, highlights the importance of identifiability in the Bayesian inference. In both cases with a constant and non-constant control intermediate variable, we use two models to estimate the PCEs under several data generating processes (DGPs). The two models seem similar in form but have different identifiability. We use the Gibbs Sampler to simulate the posterior distributions of the PCEs with 20000 iterations and the first 4000 iterations as the burn-in period. The Markov chains mix very well with the Gelman--Rubin diagnostic statistics close to one based on multiple chains. 

\subsection{Constant control intermediate variable}
\label{sec::simulation-constant}
We generate data from DGP 1:
\begin{eqnarray*}
&& Z \sim \text{Bernoulli}(0.5), \quad W \sim N(0,1), \quad Z\ind W, \quad S_1 \mid W \sim N(\gamma_0+\gamma_1W,\sigma^2),\\
&&\pr(Y_z=1\mid S_1,W) = \Phi(\beta_{z0}+\beta_{z1} S_1+\beta_{z2}W),
\end{eqnarray*}
with parameters $(\beta_{00},\beta_{01},\beta_{02})=(1,-0.5,0.5)$, $(\beta_{10},\beta_{11},\beta_{12})=(0.5,1,1.5)$ and $(\gamma_0,\gamma_1,\sigma)=(1,0.5,1)$. We name the model corresponding to DGP 1 as model 1. Under model 1, Assumption~\ref{asm:randomization} holds but the conditions in Proposition \ref{ex::nind:probit} do not. Therefore, model 1 is not identifiable. 

In DGP 2, $Z$, $W$ and $Y_z$ are the same as DGP 1, but 
 $S_1 \mid W \sim N(\gamma_0+\gamma_1W+\gamma_2W^2,\sigma^2)$, where $(\gamma_0,\gamma_1,\gamma_2,\sigma)=(1,0.5,1,1)$. We name the model corresponding to DGP 2 as model 2. Because $(1,\gamma_0+\gamma_1W+\gamma_2W^2, W)$ are linearly independent, the PCEs are identifiable based on Proposition \ref{ex::nind:probit}. 
 
 For both DGPs 1 and 2, we use the true models to analyze the generated data with sample size $1000$. We choose the following two sets of priors to assess the sensitivity of the inference based on posteriors: 
\begin{enumerate}
[(A)]
\item\label{priorA} $(\beta_{z0},\beta_{z1},\beta_{z2}) \sim \bm{N}_3(\bm{0}_3,\text{diag}(1,1,1)/10^{-2})$ for $z=0,1$, $p(\sigma^2) \propto 1/\sigma^2$, and $(\gamma_0,\gamma_1) \sim \bm{N}_2(\bm{0}_2,\text{diag}(1,1)/10^{-2})$ for model 1 (correspondingly, $(\gamma_0,\gamma_1,\gamma_2) \sim \bm{N}_3(\bm{0}_2,\text{diag}(1,1,1)/10^{-2})$ for model 2). 
 \item\label{priorB} $(\beta_{z0},\beta_{z1},\beta_{z2}) \sim \bm{N}_3(\bm{0}_3,\text{diag}(1,1,1))$ for $z=0,1$, $p(\sigma^2 )\propto 1/\sigma^2$, and $(\gamma_0,\gamma_1) \sim \bm{N}_2(\bm{0}_2,\text{diag}(1,1)/10^{-2})$ for model 1 (correspondingly, $(\gamma_0,\gamma_1,\gamma_2) \sim \bm{N}_3(\bm{0}_2,\text{diag}(1,1,1)/10^{-2})$ for model 2).
\end{enumerate}
The prior for $(\beta_{z0},\beta_{z1},\beta_{z2}) $ is much less diffused in prior \eqref{priorB} than in prior \eqref{priorA}.

\begin{figure}[ht]
 \centering
\includegraphics[width=0.8\textwidth]{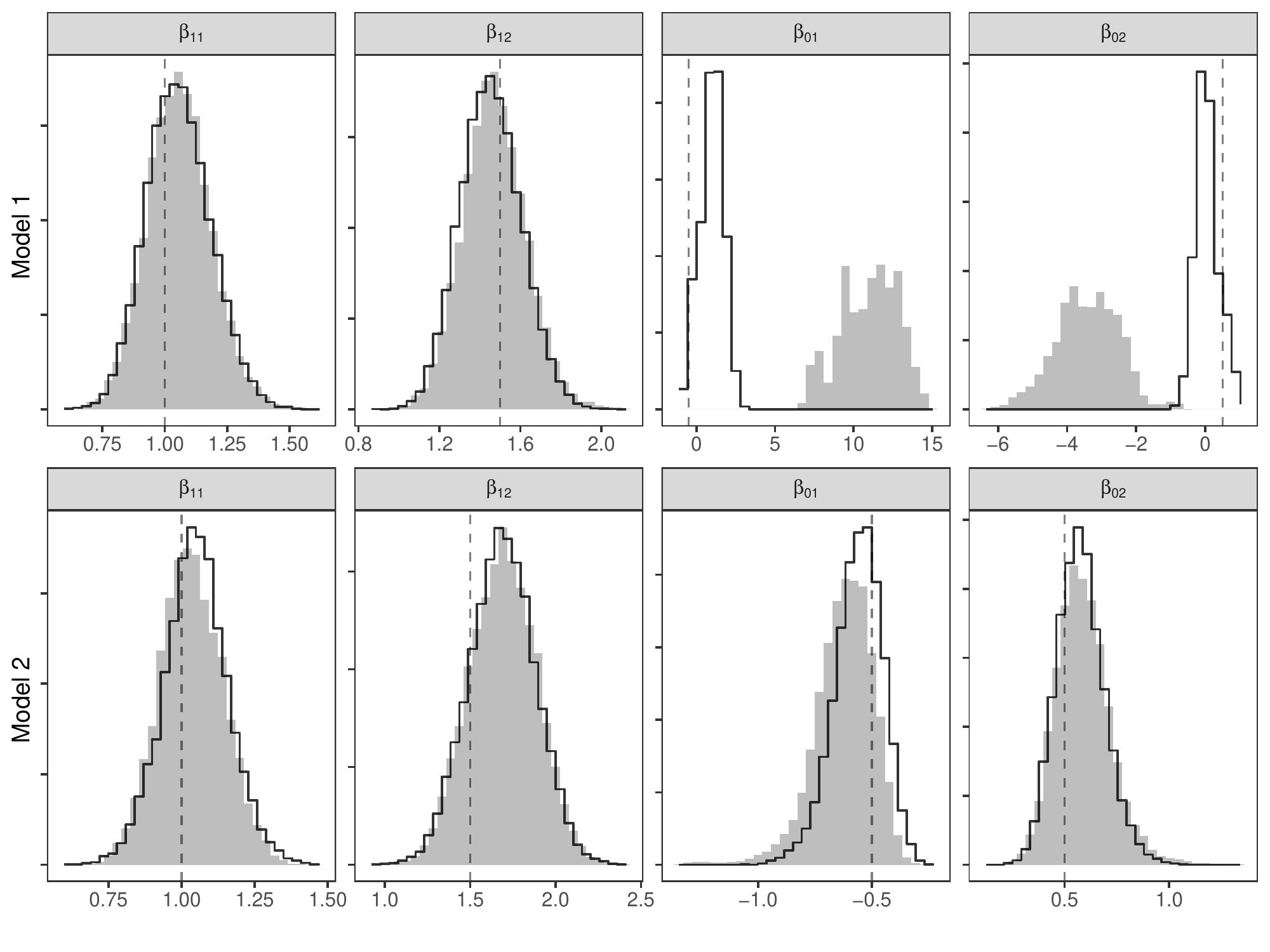}
 \caption{Posterior distributions of the parameters in Section~\ref{sec::simulation-constant}. 
 The grey histograms are the results with prior \eqref{priorA}, and the white histograms are the results with prior \eqref{priorB}. The vertical dashed lines represent the true values of the parameters.
  }
 \label{fig:YpSn} 
\end{figure}

 Figure \ref{fig:YpSn} shows the posterior distributions of $(\beta_{01},\beta_{02},\beta_{11},\beta_{12})$. For model 2, the posterior 95\% credible intervals cover the true parameters under both priors. For model 1, the posterior distributions of $\beta_{01}$ and $\beta_{02}$ differ greatly under the two priors. Their posterior distributions are far away from the true values under prior \eqref{priorA}, which shows strong evidence of non-identifiability or weakly identifiability of model 1.
 
\subsection{Non-constant control intermediate variable}
\label{sec::simulation-nonconstant}
Similar to Section 5.1, we describe two DGPs with different identifiability and evaluate the finite-sample performance of Bayesian inference under each DGP. We choose two models corresponding to two nested DGPs so that we can go beyond Section 5.1 to assess the performance of the Bayesian inference with a mis-specified model.

We first specify the two DGPs.
For DGP 3, $W \sim \text{Bernoulli}(0.5)$ and $Z \mid W=w \sim \text{Bernoulli}(\alpha_w)$, where $(\alpha_1, \alpha_2)=(0.5, 0.5)$. 
We then generate $U=(S_1,S_0)$ from
categorical distributions conditional on $W$, and $Y$ from Bernoulli distributions conditional on $Z$ and $U$ with true values of the parameters in Table~\ref{DGP3}. We name the model corresponding to DGP 3 as model 3.
For model 3, Assumptions \ref{asm:randomization} and \ref{asm:exclusion} hold. Because the stratum $(S_1,S_0)=(0,1)$ does not exist, monotonicity holds and thus the distribution of $(S_1,S_0)$ given $W$ is identifiable. From Theorem \ref{thm::nonmon:discrete}, the PCEs are identifiable. 

For DGP 4, we generate $W$ and $Z$ in the same way as DGP 4. We then generate $U=(S_1,S_0)$ from
categorical distributions conditional on $W$, and $Y$ from Bernoulli distributions conditional on $Z$ and $U$ with true values of the parameters in Table~\ref{DGP4}. We name the model corresponding to DGP 4 as model 4.
For model 4, stratum $(S_1,S_0)=(0,1)$ exists, and monotonicity does not hold. Without monotonicity, the distribution of $(S_1,S_0)\mid W$ is not identifiable, and thus the PCEs are not identifiable.

\begin{table}[ht]
\caption{True values of the parameters under DGP 3 and DGP 4.}
\label{tab::setting}
\begin{center}
\scriptsize
\subtable[DGP 3. The true PCEs are: $\ACE_{11}=0.3$, $\ACE_{10}=0.4$ and $\ACE_{00}=0.5$.]{
\begin{tabular}{ccccc}\hline
$\pr(U=u \mid W=w)$ &  $u=(1,1)$ & $u=(1,0)$ & $u=(0,0)$ &\\ \hdashline
$w=1$ &  0.5 & 0.3    & 0.2 &  \\
$w=2$ &  0.2 & 0.3 & 0.5 &\\ \hline 
$\pr(Y=1 \mid Z=z,U=u)$ &  $u=(1,1)$ & $u=(1,0)$ & $u=(0,0)$ &\\ \hdashline
$z=1$ &  0.8 & 0.7    & 0.6 &  \\
$z=0$ &  0.5 & 0.3 & 0.1 &\\ \hline 
\end{tabular}
\label{DGP3}
}

\subtable[DGP 4. The true PCEs are: $\ACE_{11}=0.3$, $\ACE_{10}=0.4$, $\ACE_{00}=0.5$ and $\ACE_{01}=-0.3$.]{
\begin{tabular}{ccccc}\hline
$\pr(U=u \mid W=w)$ &  $u=(1,1)$ & $u=(1,0)$ & $u=(0,0)$ & $u=(0,1)$\\  \hdashline
$w=1$ &  0.5 & 0.3    & 0.1 & 0.1  \\
$w=2$ &  0.1 & 0.3 & 0.5 & 0.1\\ \hline 
$\pr(Y=1 \mid Z=z,U=u)$ &  $u=(1,1)$ & $u=(1,0)$ & $u=(0,0)$ & $u=(0,1)$\\  \hdashline
$z=1$ &  0.8 & 0.7    & 0.6 & 0.2  \\
$z=0$ &  0.5 & 0.3 & 0.1 & 0.5 \\ \hline 
\end{tabular}
\label{DGP4}
}
\end{center}

\end{table}%

\paragraph{Use models 3 and 4 to analyze the data simulated from DGP 3.}
Because model 4 is a generalization of model 3, they are both correctly specified under DGP 3. However, the true value of $\ACE_{01}$ in model 4 is not well-defined. 
 
We choose two sample sizes $1000$ and $50000$. For model 3, we choose the following priors: $\pr(W=1) \sim \text{Beta}(1,1)$, $\alpha_w \sim \text{Beta}(1,1)$, and $(\pi_{11,w},\pi_{10,w},\pi_{00,w}) \sim \text{Dirichlet}(1,1,1)$ for $w=1,2$. We choose two different priors for the parameters $\delta_{u,s_1s_0}$. One is the uniform prior $\text{Beta}(1,1)$ and the other is $\text{Beta}(0.5,0.5)$. For model 4, all the priors are the same except that the prior for $(\pi_{11,w},\pi_{10,w},\pi_{00,w},\pi_{01,w})$ is $ \text{Dirichlet}(1,1,1,1)$.

\begin{figure}
 \centering
\subfigure{
  \label{fig:YbSb1} 
\includegraphics[width=0.8\textwidth]{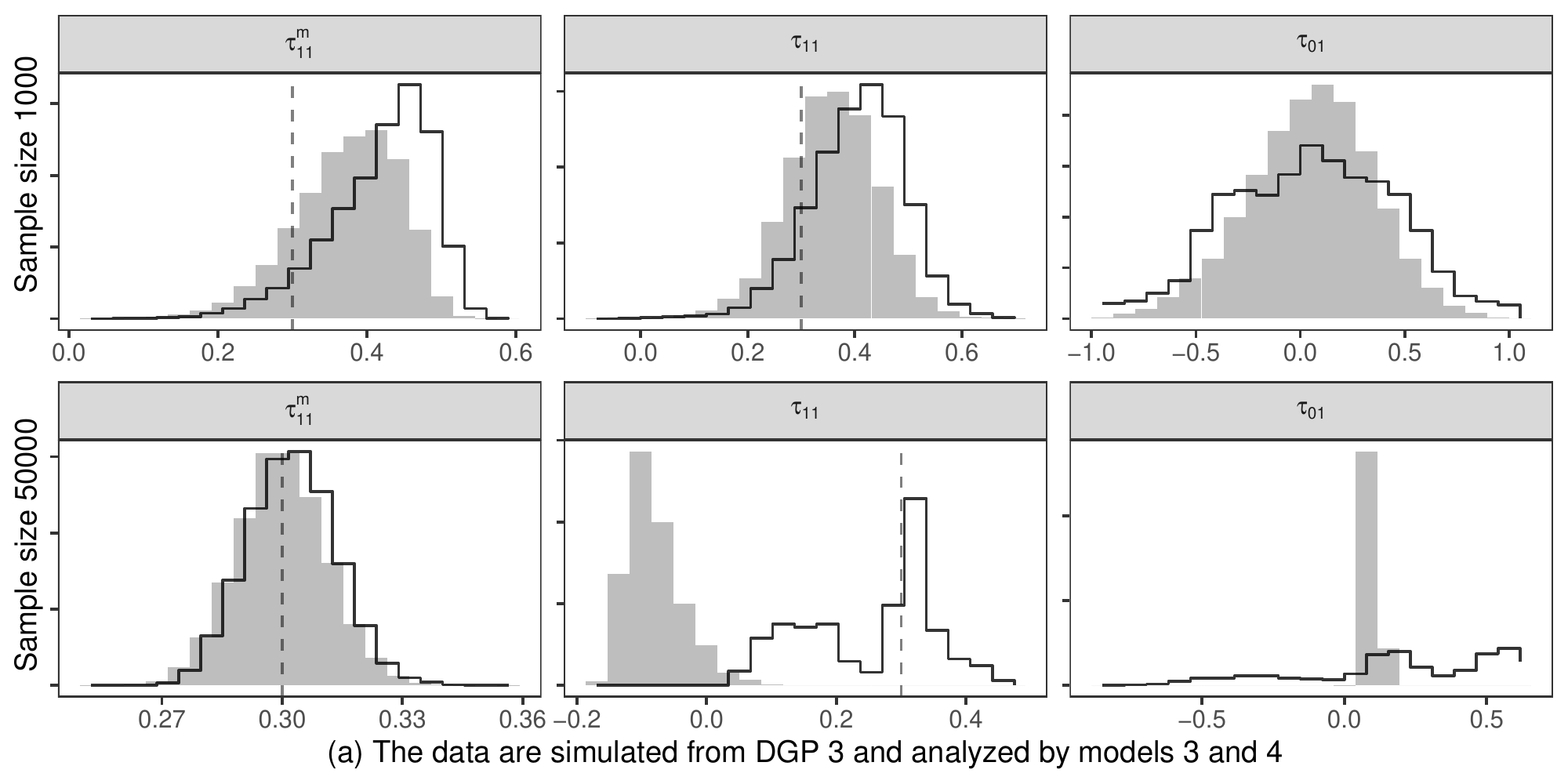}}
\subfigure{
  \label{fig:YbSb2} 
\includegraphics[width=0.8\textwidth]{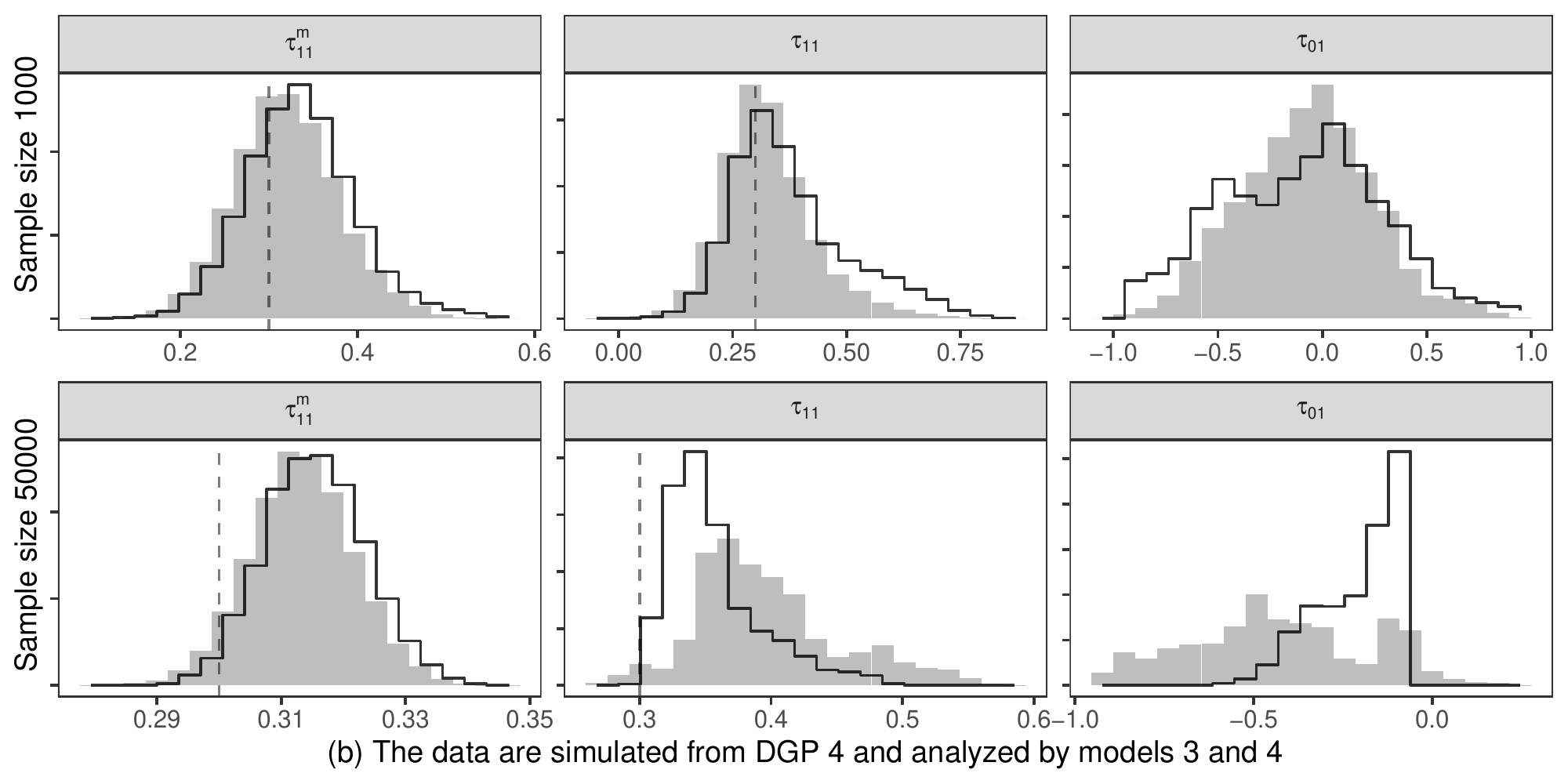}}
 \caption{Posterior distributions of the PCEs in Section~\ref{sec::simulation-nonconstant}. $\tau^{\text{m}}_{11}$ is the PCE within the stratum $(S_1,S_0)=(1,1)$ under model 3; $\tau_{11}$ and $\tau_{01}$ are the PCEs within the strata $(S_1,S_0)=(1,1)$ and $(0,1)$ under model 4. The grey histograms are the results with prior $\text{Beta}(1,1)$ for $\delta_{u,s_1s_0}$, and the white histograms are the results with prior $\text{Beta}(0.5,0.5)$ for $\delta_{u,s_1s_0}$.
  The vertical dashed lines represent the true values of the parameters. 
  }
 \label{fig:YbSb}  
\end{figure}

Figure \ref{fig:YbSb}(a) shows the posterior distributions of $\ACE^{\text{m}}_{11}$, $\ACE_{11}$ and $\ACE_{01}$, where $\ACE^{\text{m}}_{11}$ is the PCE within the stratum $(S_1,S_0)=(1,1)$ under model 3, and $\ACE_{11}$ and $\ACE_{01}$ are the PCEs within the strata $(S_1,S_0)=(1,1)$ and $(0,1)$ under model 4, respectively. Comparing the two rows of plots in Figure \ref{fig:YbSb1}, we can see that as the sample size increases, the posterior 95\% credible intervals of $\ACE^{\text{m}}_{11}$ becomes narrower and always cover the true value, regardless of the priors. For model 4, the posterior distributions of the PCEs change greatly and the posterior 95\% credible intervals do not shrink as those under model 3. When the sample size is $50000$, the posterior distribution of $\ACE_{11}$ is far away from the true value with the flat prior $\text{Beta}(1,1)$ and  is not unimodal with the prior $\text{Beta}(0.5,0.5)$. This is in sharp contrast to standard Bayesian problems in which the Beta$(1,1)$ and Beta$(0.5,0.5)$ priors result in small discrepancies. The drastic differences with different sample sizes and priors show strong evidence of the non-identifiability or weakly identifiability of model 4, which can yield misleading estimates and inferences.

\paragraph{Use models 3 and 4 to analyze data simulated from DGP 4.} 
The true model 4 is not identifiable, and model 3 is mis-specified. Figure \ref{fig:YbSb}(b) shows the results for $\ACE^{\text{m}}_{11}$, $\ACE_{11}$ and $\ACE_{01}$. Although model 3 is not the true model, the result under this model is very stable under different priors. The 95\% credible intervals of $\ACE^{\text{m}}_{11}$ cover the true value. This may be due to our choice of small values of $\pi_{01,1}$ and $\pi_{01,2}$, which makes model 3 only slightly deviates from the true model. In contrast, the result of model 4 changes greatly under different priors even when the sample size is large. The posterior distributions of $\ACE_{01}$ are multimodal even with a very large sample size.  Therefore, using an unidentifiable model may lead to an undesirable result even if it is a true model.

Our simulation demonstrates that identification is important in the Bayesian inference. Otherwise, the results are extremely sensitive to the priors. 
More importantly, the simulation suggests that when the proposed model is not identifiable, using an identifiable model ``close'' to it may be a compromising solution.

\section{Application to the Job Search Intervention Study}\label{sec::application}

The Job Search Intervention Study was a randomized field experiment investigating the efficacy of a job training intervention on unemployed workers \citep{vinokur1995impact, vinokur1997mastery,tingley2014mediation}.
The program was designed not only to increase reemployment among the unemployed but also to enhance the mental health of the job seekers.  
In the study, 600 unemployed workers are randomly assigned to the treatment group ($Z=1$) and 299 are assigned to the control group ($Z=0$). Those in the treatment group participated in workshops that covered skills for job search and coping with stress. Those in the control group received a booklet describing job-search tips. The intermediate variable $S$ is a measure of job-search self-efficacy ranged from $1$ to $5$. It measures the participants' confidence in being able to successfully perform six essential job-search activities including completing a job application or resume, using their social network to discover promising job openings, and getting their point across in a job interview. The outcome $Y$ is a measure of depressive symptoms based on the Hopkins Symptom Checklist. 
It measures how much they had been bothered or distressed in the last two weeks by various depression symptoms such as feeling blue, having thoughts of ending one's life, and crying easily.
Let $W$ be the previous occupation, which is a nominal variable with seven categories.

We assume that $(S_1,S_0) $ given $W$ follows \eqref{eq::bivariateNormal}, where $\rho(w)$ is the correlation coefficient of $S_1$ and $S_0$ given $W=w$. We further assume linear models for $Y_1$ and $Y_0$:
\begin{eqnarray*}
&&Y_z= \beta_{z0}+\beta_{z1} S_1+\beta_{z2} S_0+\epsilon_{Y_z},
\end{eqnarray*}
where $\epsilon_{Y_1} \sim N(0,\sigma_{Y_1}^2)$, $\epsilon_{Y_0} \sim N(0,\sigma_{Y_0}^2)$, 
 and $(\epsilon_{Y_1},\epsilon_{Y_0}) \ind (S_1,S_0,W)$. 
 We choose the linear model because of its simplicity for illustration, and acknowledge its limitation and leave the task of building more flexible models for $Y_1$ and $Y_0$ to future work. 
Under this model, the PCEs equals 
\begin{eqnarray*}
\tau_{s_1 s_0}= \beta_{10}- \beta_{00}+(\beta_{11}-\beta_{01}) s_1+(\beta_{12}-\beta_{02}) s_0.
\end{eqnarray*}
We assume $\rho(w) = \rho$ and treat $\rho$ as the sensitivity parameter within $\{ 0,0.2,0.4,0.6,0.8 \}$. 
From Proposition~\ref{cor:trivariateN}, the PCEs are identifiable. We use a Bayesian approach and simulate the posterior distributions of the PCEs. 
 To assess the sensitivity of our results to different priors, we choose two different priors. 
 Denote $\bm{\beta}_1=(\beta_{10},\beta_{11},\beta_{12})$, $\bm{\beta}_0=(\beta_{00},\beta_{01},\beta_{02})$ and $\bm{\mu}_{w} =(\mu_1(w),\mu_0(w))$. 
For the first prior, we choose multivariate Normal priors for $\bm{\beta}_z$ and $\bm{\mu}_{w}$: $\bm{\beta}_z \sim \bm{N}_3(\bm{0},\bOmega_z)$, $\bm{\mu}_{w} \sim \bm{N}_2(\bm{0},\bOmega)$, with $\bOmega_z=10^2\text{diag}(1,1,1)$ and $\bOmega=10^2\text{diag}(1,1)$ for $z=0,1$, and $w=1,\dots,7$. We choose the following non-informative parameters for the other parameter: $f(\sigma^2_{zw}) \propto 1/ \sigma^2_{zw}$, $f(\sigma^2_{Y_z}) \propto 1/\sigma^2_{Y_z}$, $\{\pr(W=1),\ldots,\pr(W=7)\} \sim \text{Dirichlet}(1,\ldots,1)$ and $\pr(Z=1\mid W=w) \sim \text{Beta}(1,1)$, where $z=0,1$ and $w=1,\dots,7$. For the second prior, we choose $\bOmega_z=\text{diag}(1,1,1)$ and $\bOmega=\text{diag}(1,1)$ and keep other prior distributions unchanged. 
We will present the results for the first prior in the main text and show the sensitivity check of the results to different priors in Appendix C.2.

\begin{figure}[t]
 \centering
\includegraphics[width=0.6\textwidth]{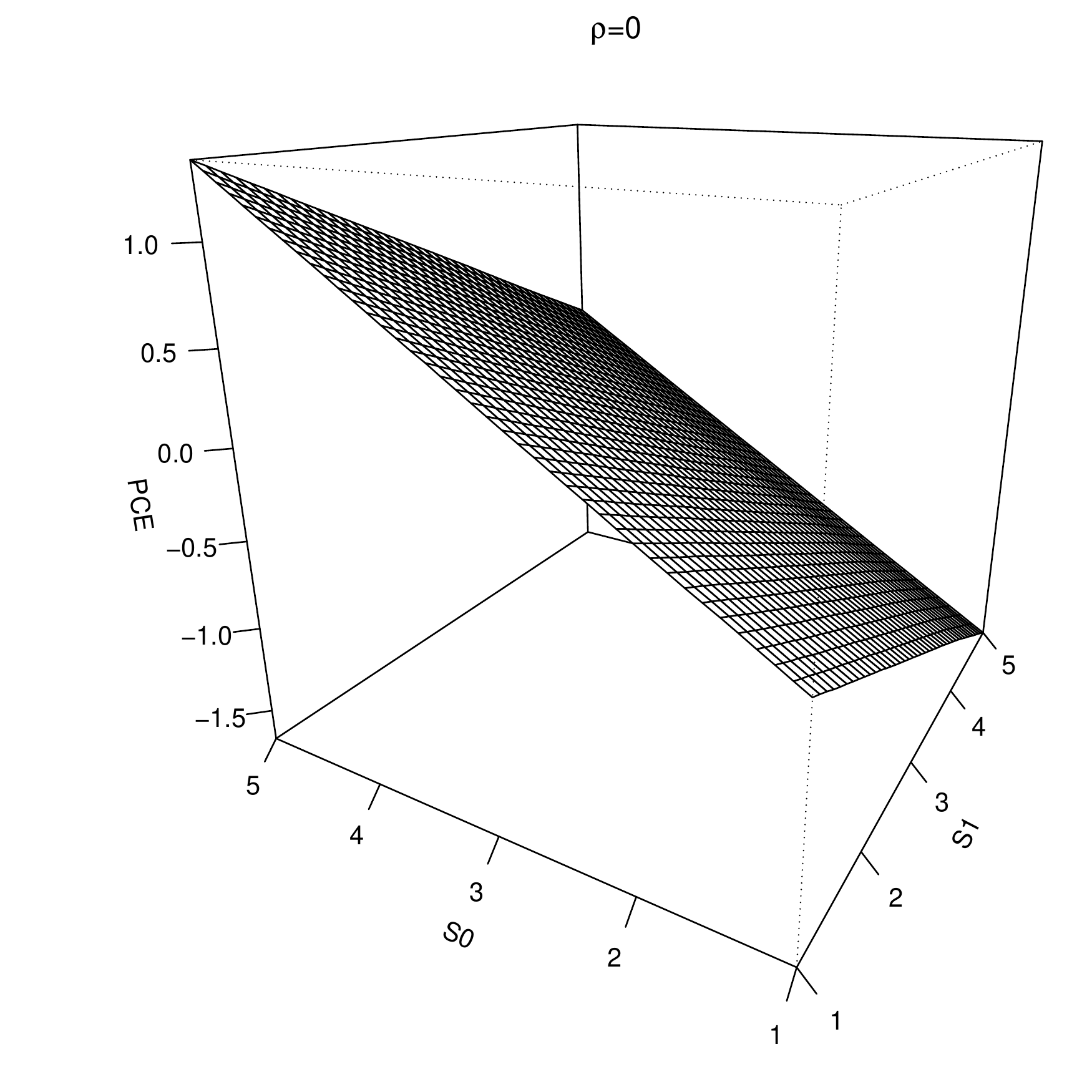}
 \caption{Posterior medians of the PCEs with $\rho= 0$.}

 \label{fig:jobs} 
\end{figure}

\begin{table}[htp]
\caption{Point and interval estimates of some PCEs using the Bayesian approach. The intervals excluding zero are highlighted in bold.}
\label{tab:app}
\begin{center}
\scriptsize
\begin{tabular}{cccccc}

\hline
    $(S_1,S_0) $           &  $\rho$=0 &  $\rho$=0.2  & $\rho$=0.4  &$\rho$= 0.6  &  $\rho$=0.8  \\
               \cline{2-6}
$(1.00,5.00)$ & 1.363 & 1.901 & 1.676 & 1.790 & 1.530\\
           & $(-0.332,3.164)$ & $(-0.504,4.681)$ & $(-0.837,4.125)$ & $(-1.167,5.182)$ &$(-1.331,4.832)$\\
 $(3.67,4.50)$ & $0.288$ & $0.392$ & $0.366$ & $0.389$ & $0.318$\\
           & $(-0.009,0.613)$ & $(-0.053,0.962)$ & $(-0.143,0.876)$ & $(-0.240,1.107)$ &$(-0.310,1.047)$\\
 $(4.17,4.00)$ & $-0.093$ & $-0.112$ & $-0.100$ & $-0.104$ & $-0.099$\\
           & $(-0.197,0.009)$ & $(-0.227,0.004)$ & $(-0.220,0.009)$ & $(-0.240,0.011)$ &$(-0.234,0.017)$\\
  $(4.67,3.58)$ & $-0.439$ & $-0.563$ & $-0.522$ & $-0.550$ & $-0.476$\\
          & $\bm{(-0.815,-0.077)}$ & $\bm{(-1.202,-0.030)}$ & $(-1.104,0.053)$ & $(-1.362,0.152)$ &$(-1.315,0.230)$\\
  $(5.00,1.67)$ & $-1.386$ & $-1.732$ & $-1.700$ & $-1.773$ & $-1.496$\\
           & $\bm{(-2.451,-0.428)}$ & $\bm{(-3.428,-0.338)}$ & $\bm{(-3.451,-0.011)}$ & $(-4.251,0.368)$ &$(-4.166,0.717)$\\
            \hline

\label{tab:sens}

\end{tabular}

\end{center}

\end{table}%

Figure \ref{fig:jobs} shows the posterior medians of $\tau_{s_1 s_0}$ for all $(s_1,s_0)$ under $\rho=0$. The surface of these posterior medians rises from its lowest point at principal stratum $(5,1)$ to its highest point at principal stratum $(1,5)$. In general, the estimated PCE increases as the difference between $S_1$ and $S_0$ decreases. That is, for people who can gain more for the job-search self-efficacy from the treatment, the treatment can lower the risk of depression to a larger extent. 
\citet{imai2010general} analyzed this data using a mediation analysis and found that the indirect effect of the treatment through job-search self-efficacy is negative. This implies the program participation decreases depressive symptoms by increasing the level of job search self-efficacy. 
\citet{jo2011use} used the principal stratification approach by dichotomizing the job-search self-efficacy, and found that the treatment has a negative effect on the depression for people whose job-search self-efficacy is improved by the treatment. Our conclusion corroborates with their findings.

In our analysis, we assume Assumption~\ref{asm:exclusion} holds conditional on the previous occupation. 
It is plausible because conditional on the potential values of the job-search self-efficacy, the previous occupation may not affect the depressive symptoms. In Appendix C.3, we also conduct an analysis of Assumption~\ref{asm:exclusion} by including more covariates.
  
To assess the sensitivity of the PCEs to $\rho$, we choose five principal strata, consisting of the maximum, minimum, 25\%, 50\%, and 75\% quantiles of $S_1$ and $S_0$. Table~\ref{tab:sens} shows their posterior medians and 95\% credible intervals.
The point estimates are not sensitive to the values of $\rho$, and the interval estimates are not sensitive to small values of $\rho$. But as $\rho$ grows larger, the intervals tend to become wider which makes the results not significant.

Two technical issues arise in our data analysis. First, Proposition~\ref{cor:trivariateN} requires $W$ to be continuous but $W$ is categorical in our application. In Appendix C.1, we give a formal justification of the identifiability of the PCEs in our model with a discrete $W$. Second, the Normality assumptions on the outcomes are invoked for convenience in the Bayesian computation. In fact, without Normality, we can use the method of moments to estimate the PCEs. The results from the method of moments are similar to those from the Bayesian inference; see Appendix C.2.

\section{Discussion}\label{sec::discussion}

\subsection{Summary and extensions}

Identification of the PCEs is an important but challenging problem. Although several empirical studies have leveraged auxiliary variables to improve inference for the PCEs, formal identification results have not been established especially for non-binary intermediate variables. Our results supplement previous empirical studies with theoretical justifications for identification. 
We give identification results for several models based on Normal distributions, which can be generalized to other commonly-used distributions.  Appendix B.4 gives identification results for models based on $t$ distributions, which are useful for robust analysis of data with heavy tails.


Researchers have conducted sensitivity analyses for the principal ignorability and the auxiliary independence using different kinds of models under various settings. For example, \citet{ding2017principal} proposed the sensitivity analysis for principal ignorability with a binary intermediate variable, and \citet{jiang2016principal} proposed the sensitivity analysis for auxiliary independence using a random-effects model. However, there is no general setup for the sensitivity analysis of these assumptions, which depends on the specification of the model and types of the outcomes and the intermediate variables. We believe that sensitivity analysis should be routinely conducted in problems with principal stratification, but leave the development and the technical details to future research.

\subsection{Comparing two strategies}

Auxiliary variables play different roles in identifying the PCEs, depending on the underlying assumptions. Under principal ignorability, auxiliary variables can be viewed as ``confounders'' between the principal stratification variable and the outcome.
In contrast, under auxiliary independence, auxiliary variables can be treated as an ``instrumental variables'' for the relationship between the principal stratification and the outcome.  Therefore, the comparison between the principal ignorability and auxiliary independence for identifying the PCEs resembles the comparison between the ignorability assumption \citep{rosenbaum1983central} and the instrumental variable method \citep{angrist1996identification} for identifying the average causal effect. The methods based on principal ignorability are easy to employ because the assumption generally conditions on all baseline variables. However, they bear similar disadvantages as the methods based on ignorability for estimating average causal effect --- we do not know whether we have conditioned on sufficient variables \citep{judea2000causality,pearl2009letter}. 
In contrast, the methods based on auxiliary independence may be burdening to analysts and content experts because one needs to carve out a specific baseline variable as a designated auxiliary variable. However, the advantage is that we can intentionally target the variable based on science and experts' knowledge or by design. 
For example, this assumption can possibly be used in a multi-center trial as in Example~\ref{ex::jiang}, and in the augmented design for assessing the effect of vaccination  as in Example~\ref{ex::follmann}.

Although we restrict the auxiliary variable $W$ to be pretreatment in the paper, the auxiliary independence assumption allows for it to be affected by the treatment. It only requires the auxiliary variable to be independent of the outcome conditional on the treatment and principal strata, which can hold even if the auxiliary variable is posttreatment. For example,
for a binary $S$, \citet{mealli2013using} identify the PCEs in completely randomized experiments using a secondary outcome as the auxiliary variable. In contrast, the principal ignorability assumption is unlikely to hold with a posttreatment auxiliary variable. The required independence would fail due to the bias induced by conditioning on a posttreatment variable.

%
%
%

\subsection{Alternative identification strategies}

Alternative identification strategies do exist without requiring an auxiliary variable. For a binary intermediate variable, without monotonicity or exclusion restriction, \citet{hirano2000assessing} suggested using parallel outcome models to improve identifiability where the regression coefficients of the covariates are the same for all types of non-compliers. \citet{mealli2016identification} used the concentration graph theory to study the identification of the PCEs. It is of interest to combine these strategies in theory and practice.

The identification issue of PCEs is closely related to the finite mixture models. For example, with a binary intermediate variable, the observed data with $(Z=1,S=1)$ is a mixture of principal strata $(S_1=1,S_0=1)$ and $(S_1=1,S_0=0)$, and the observed data with $(Z=1,S=0)$ is a mixture of principal strata $(S_1=0,S_0=0)$ and $(S_1=0,S_0=1)$.  
From this perspective, principal ignorability and auxiliary independence help to separate the components in the finite mixture model. 
Researchers sometimes use parametric finite mixture models for principal stratification problems \citep{zhang2009likelihood, frumento2012evaluating}. 
However, even if those models are parametrically identifiable, the estimators often have poor finite-sample properties \citep{frumento2016fragility, feller2016weak}.
These findings echo the caveat from \citet[][page 96]{cox2011principles}: ``If an issue can be addressed nonparametrically then it will often be
 better to tackle it parametrically; however, if it cannot be
 resolved nonparametrically then it is usually dangerous to resolve
 it parametrically.''
 This is an important motivation for us to seek for nonparametric and semiparametric identifiability as presented in this paper.


\bibliographystyle{Chicago}

\bibliography{PSACEidentification}

\begin{thebibliography}{}

\bibitem[\protect\citeauthoryear{Abramowitz, Stegun, et~al.}{Abramowitz
  et~al.}{1966}]{abramowitz1966handbook}
Abramowitz, M., I.~A. Stegun, et~al. (1966).
\newblock Handbook of mathematical functions.
\newblock {\em Applied Mathematics Series\/}~{\em 55}, 62.

\bibitem[\protect\citeauthoryear{Angrist, Imbens, and Rubin}{Angrist
  et~al.}{1996}]{angrist1996identification}
Angrist, J.~D., G.~W. Imbens, and D.~B. Rubin (1996).
\newblock Identification of causal effects using instrumental variables (with
  discussion).
\newblock {\em Journal of the American Statistical Association\/}~{\em 91},
  444--455.

\bibitem[\protect\citeauthoryear{Bartolucci and Grilli}{Bartolucci and
  Grilli}{2011}]{bartolucci2011modeling}
Bartolucci, F. and L.~Grilli (2011).
\newblock Modeling partial compliance through copulas in a principal
  stratification framework.
\newblock {\em Journal of the American Statistical Association\/}~{\em 106},
  469--479.

\bibitem[\protect\citeauthoryear{Bickel and Doksum}{Bickel and
  Doksum}{2015}]{bickel2015mathematical}
Bickel, P.~J. and K.~A. Doksum (2015).
\newblock {\em Mathematical Statistics: Basic Ideas and Selected Topics, Volume
  I}.
\newblock Boca Raton, FL: CRC Press.

\bibitem[\protect\citeauthoryear{Cheng and Small}{Cheng and
  Small}{2006}]{cheng2006bounds}
Cheng, J. and D.~S. Small (2006).
\newblock Bounds on causal effects in three-arm trials with non-compliance.
\newblock {\em Journal of the Royal Statistical Society: Series B (Statistical
  Methodology)\/}~{\em 68}, 815--836.

\bibitem[\protect\citeauthoryear{Cochran}{Cochran}{1957}]{cochran1957analysis}
Cochran, W.~G. (1957).
\newblock Analysis of covariance: its nature and uses.
\newblock {\em Biometrics\/}~{\em 13}, 261--281.

\bibitem[\protect\citeauthoryear{Conlon, Taylor, and Elliott}{Conlon
  et~al.}{2017}]{conlon2014surrogacy}
Conlon, A., J.~Taylor, and M.~Elliott (2017).
\newblock Surrogacy assessment using principal stratification and a {G}aussian
  copula model.
\newblock {\em Statistical Methods in Medical Research\/}~{\em 26}, 88--107.

\bibitem[\protect\citeauthoryear{Cox and Donnelly}{Cox and
  Donnelly}{2011}]{cox2011principles}
Cox, D.~R. and C.~A. Donnelly (2011).
\newblock {\em Principles of applied statistics}.
\newblock Cambridge: Cambridge University Press.

\bibitem[\protect\citeauthoryear{Daniels, Roy, Kim, Hogan, and Perri}{Daniels
  et~al.}{2012}]{daniels2012bayesian}
Daniels, M.~J., J.~A. Roy, C.~Kim, J.~W. Hogan, and M.~G. Perri (2012).
\newblock Bayesian inference for the causal effect of mediation.
\newblock {\em Biometrics\/}~{\em 68}, 1028--1036.

\bibitem[\protect\citeauthoryear{Ding}{Ding}{2016}]{ding2016conditional}
Ding, P. (2016).
\newblock On the conditional distribution of the multivariate $t$ distribution.
\newblock {\em The American Statistician\/}~{\em 70}, 293--295.

\bibitem[\protect\citeauthoryear{Ding, Geng, Yan, and Zhou}{Ding
  et~al.}{2011}]{ding2011identifiability}
Ding, P., Z.~Geng, W.~Yan, and X.~H. Zhou (2011).
\newblock Identifiability and estimation of causal effects by principal
  stratification with outcomes truncated by death.
\newblock {\em Journal of the American Statistical Association\/}~{\em 106},
  1578--1591.

\bibitem[\protect\citeauthoryear{Ding and Li}{Ding and
  Li}{2018}]{ding2018causal}
Ding, P. and F.~Li (2018).
\newblock Causal inference: A missing data perspective.
\newblock {\em Statistical Science\/}~{\em 33}, 214--237.

\bibitem[\protect\citeauthoryear{Ding and Lu}{Ding and
  Lu}{2017}]{ding2017principal}
Ding, P. and J.~Lu (2017).
\newblock Principal stratification analysis using principal scores.
\newblock {\em Journal of the Royal Statistical Society: Series B (Statistical
  Methodology)\/}~{\em 79}, 757--777.

\bibitem[\protect\citeauthoryear{Efron and Feldman}{Efron and
  Feldman}{1991}]{efron1991compliance}
Efron, B. and D.~Feldman (1991).
\newblock Compliance as an explanatory variable in clinical trials (with
  discussion).
\newblock {\em Journal of the American Statistical Association\/}~{\em 86},
  9--17.

\bibitem[\protect\citeauthoryear{Elliott, Raghunathan, and Li}{Elliott
  et~al.}{2010}]{elliott2010bayesian}
Elliott, M.~R., T.~E. Raghunathan, and Y.~Li (2010).
\newblock Bayesian inference for causal mediation effects using principal
  stratification with dichotomous mediators and outcomes.
\newblock {\em Biostatistics\/}~{\em 11}, 353--372.

\bibitem[\protect\citeauthoryear{Everitt and Hand}{Everitt and
  Hand}{1981}]{everitt1981finite}
Everitt, B.~S. and D.~J. Hand (1981).
\newblock {\em Finite Mixture Distributions}.
\newblock New York: Chapman and Hall.

\bibitem[\protect\citeauthoryear{Feller, Greif, Ho, Miratrix, and
  Pillai}{Feller et~al.}{2019}]{feller2016weak}
Feller, A., E.~Greif, N.~Ho, L.~Miratrix, and N.~Pillai (2019).
\newblock Weak separation in mixture models and implications for principal
  stratification.
\newblock {\em arXiv preprint arXiv:1602.06595\/}.

\bibitem[\protect\citeauthoryear{Follmann}{Follmann}{2006}]{follmann2006augmented}
Follmann, D. (2006).
\newblock Augmented designs to assess immune response in vaccine trials.
\newblock {\em Biometrics\/}~{\em 62}, 1161--1169.

\bibitem[\protect\citeauthoryear{Follmann}{Follmann}{2000}]{follmann2000effect}
Follmann, D.~A. (2000).
\newblock On the effect of treatment among would-be treatment compliers: An
  analysis of the multiple risk factor intervention trial.
\newblock {\em Journal of the American Statistical Association\/}~{\em 95},
  1101--1109.

\bibitem[\protect\citeauthoryear{Forastiere, Mattei, and Ding}{Forastiere
  et~al.}{2018}]{forastiere2018principal}
Forastiere, L., A.~Mattei, and P.~Ding (2018).
\newblock Principal ignorability in mediation analysis: through and beyond
  sequential ignorability.
\newblock {\em Biometrika\/}~{\em 105}, 979--986.

\bibitem[\protect\citeauthoryear{Frangakis and Rubin}{Frangakis and
  Rubin}{2002}]{frangakis2002principal}
Frangakis, C.~E. and D.~B. Rubin (2002).
\newblock Principal stratification in causal inference.
\newblock {\em Biometrics\/}~{\em 58}, 21--29.

\bibitem[\protect\citeauthoryear{Frumento, Mealli, Pacini, and Rubin}{Frumento
  et~al.}{2012}]{frumento2012evaluating}
Frumento, P., F.~Mealli, B.~Pacini, and D.~B. Rubin (2012).
\newblock Evaluating the effect of training on wages in the presence of
  noncompliance, nonemployment, and missing outcome data.
\newblock {\em Journal of the American Statistical Association\/}~{\em 107},
  450--466.

\bibitem[\protect\citeauthoryear{Frumento, Mealli, Pacini, and Rubin}{Frumento
  et~al.}{2016}]{frumento2016fragility}
Frumento, P., F.~Mealli, B.~Pacini, and D.~B. Rubin (2016).
\newblock The fragility of standard inferential approaches in principal
  stratification models relative to direct likelihood approaches.
\newblock {\em Statistical Analysis and Data Mining\/}~{\em 9}, 58--70.

\bibitem[\protect\citeauthoryear{Gabriel and Follmann}{Gabriel and
  Follmann}{2016}]{gabriel2016augmented}
Gabriel, E.~E. and D.~Follmann (2016).
\newblock Augmented trial designs for evaluation of principal surrogates.
\newblock {\em Biostatistics\/}~{\em 17}, 453--467.

\bibitem[\protect\citeauthoryear{Gallop, Small, Lin, Elliott, Joffe, and
  Ten~Have}{Gallop et~al.}{2009}]{gallop2009mediation}
Gallop, R., D.~S. Small, J.~Y. Lin, M.~R. Elliott, M.~Joffe, and T.~R. Ten~Have
  (2009).
\newblock Mediation analysis with principal stratification.
\newblock {\em Statistics in Medicine\/}~{\em 28}, 1108--1130.

\bibitem[\protect\citeauthoryear{Gelman, Carlin, Stern, Dunson, Vehtari, and
  Rubin}{Gelman et~al.}{2014}]{gelman2014bayesian}
Gelman, A., J.~B. Carlin, H.~S. Stern, D.~B. Dunson, A.~Vehtari, and D.~B.
  Rubin (2014).
\newblock {\em Bayesian Data Analysis (3rd ed.)}.
\newblock London: Chapman and Hall/CRC.

\bibitem[\protect\citeauthoryear{Gilbert and Hudgens}{Gilbert and
  Hudgens}{2008}]{gilbert2008evaluating}
Gilbert, P.~B. and M.~G. Hudgens (2008).
\newblock Evaluating candidate principal surrogate endpoints.
\newblock {\em Biometrics\/}~{\em 64}, 1146--1154.

\bibitem[\protect\citeauthoryear{Gustafson}{Gustafson}{2009}]{gustafson2009limits}
Gustafson, P. (2009).
\newblock What are the limits of posterior distributions arising from
  nonidentified models, and why should we care?
\newblock {\em Journal of the American Statistical Association\/}~{\em 104},
  1682--1695.

\bibitem[\protect\citeauthoryear{Gustafson}{Gustafson}{2015}]{gustafson2015bayesian}
Gustafson, P. (2015).
\newblock {\em Bayesian Inference for Partially Identified Models: Exploring
  the Limits of Limited Data}.
\newblock Chapman and Hall/CRC.

\bibitem[\protect\citeauthoryear{Hirano, Imbens, Rubin, and Zhou}{Hirano
  et~al.}{2000}]{hirano2000assessing}
Hirano, K., G.~W. Imbens, D.~B. Rubin, and X.-H. Zhou (2000).
\newblock Assessing the effect of an influenza vaccine in an encouragement
  design.
\newblock {\em Biostatistics\/}~{\em 1}, 69--88.

\bibitem[\protect\citeauthoryear{Hu and Shiu}{Hu and
  Shiu}{2017}]{hu2017nonparametric}
Hu, Y. and J.-L. Shiu (2017).
\newblock Nonparametric identification using instrumental variables: sufficient
  conditions for completeness.
\newblock {\em Econometric Theory\/}~{\em 34}, 1--35.

\bibitem[\protect\citeauthoryear{Huang and Gilbert}{Huang and
  Gilbert}{2011}]{huang2011comparing}
Huang, Y. and P.~B. Gilbert (2011).
\newblock Comparing biomarkers as principal surrogate endpoints.
\newblock {\em Biometrics\/}~{\em 67}, 1442--1451.

\bibitem[\protect\citeauthoryear{Hudgens and Gilbert}{Hudgens and
  Gilbert}{2009}]{hudgens2009assessing}
Hudgens, M.~G. and P.~B. Gilbert (2009).
\newblock Assessing vaccine effects in repeated low-dose challenge experiments.
\newblock {\em Biometrics\/}~{\em 65}, 1223--1232.

\bibitem[\protect\citeauthoryear{Imai}{Imai}{2008}]{imai2008sharp}
Imai, K. (2008).
\newblock Sharp bounds on the causal effects in randomized experiments with
  ``truncation-by-death''.
\newblock {\em Statistics and Probability Letters\/}~{\em 78}, 144--149.

\bibitem[\protect\citeauthoryear{Imai, Keele, and Tingley}{Imai
  et~al.}{2010}]{imai2010general}
Imai, K., L.~Keele, and D.~Tingley (2010).
\newblock A general approach to causal mediation analysis.
\newblock {\em Psychological Methods\/}~{\em 15}, 309--334.

\bibitem[\protect\citeauthoryear{Imbens and Rubin}{Imbens and
  Rubin}{2015}]{imbens2015causal}
Imbens, G.~W. and D.~B. Rubin (2015).
\newblock {\em Causal Inference in Statistics, Social, and Biomedical
  Sciences}.
\newblock Cambridge: Cambridge University Press.

\bibitem[\protect\citeauthoryear{Jiang, Ding, and Geng}{Jiang
  et~al.}{2016}]{jiang2016principal}
Jiang, Z., P.~Ding, and Z.~Geng (2016).
\newblock Principal causal effect identification and surrogate end point
  evaluation by multiple trials.
\newblock {\em Journal of the Royal Statistical Society: Series B (Statistical
  Methodology)\/}~{\em 78}, 829--848.

\bibitem[\protect\citeauthoryear{Jin and Rubin}{Jin and
  Rubin}{2008}]{jin2008principal}
Jin, H. and D.~B. Rubin (2008).
\newblock Principal stratification for causal inference with extended partial
  compliance.
\newblock {\em Journal of the American Statistical Association\/}~{\em 103},
  101--111.

\bibitem[\protect\citeauthoryear{Jo}{Jo}{2008}]{jo2008causal}
Jo, B. (2008).
\newblock Causal inference in randomized experiments with mediational
  processes.
\newblock {\em Psychological Methods\/}~{\em 13}, 314--336.

\bibitem[\protect\citeauthoryear{Jo and Stuart}{Jo and
  Stuart}{2009}]{jo2009use}
Jo, B. and E.~A. Stuart (2009).
\newblock On the use of propensity scores in principal causal effect
  estimation.
\newblock {\em Statistics in Medicine\/}~{\em 28}, 2857--2875.

\bibitem[\protect\citeauthoryear{Jo, Stuart, MacKinnon, and Vinokur}{Jo
  et~al.}{2011}]{jo2011use}
Jo, B., E.~A. Stuart, D.~P. MacKinnon, and A.~D. Vinokur (2011).
\newblock The use of propensity scores in mediation analysis.
\newblock {\em Multivariate Behavioral Research\/}~{\em 46}, 425--452.

\bibitem[\protect\citeauthoryear{Joffe, Small, Hsu, et~al.}{Joffe
  et~al.}{2007}]{joffe2007defining}
Joffe, M.~M., D.~Small, C.-Y. Hsu, et~al. (2007).
\newblock Defining and estimating intervention effects for groups that will
  develop an auxiliary outcome.
\newblock {\em Statistical Science\/}~{\em 22}, 74--97.

\bibitem[\protect\citeauthoryear{Kim, Henneman, Choirat, and Zigler}{Kim
  et~al.}{2020}]{kim2020health}
Kim, C., L.~R. Henneman, C.~Choirat, and C.~M. Zigler (2020).
\newblock Health effects of power plant emissions through ambient air quality.
\newblock {\em Journal of the Royal Statistical Society: Series A (Statistics
  in Society)\/}.

\bibitem[\protect\citeauthoryear{Lange, Little, and Taylor}{Lange
  et~al.}{1989}]{lange1989robust}
Lange, K.~L., R.~J. Little, and J.~M. Taylor (1989).
\newblock Robust statistical modeling using the $t$ distribution.
\newblock {\em Journal of the American Statistical Association\/}~{\em 84},
  881--896.

\bibitem[\protect\citeauthoryear{Lehmann and Romano}{Lehmann and
  Romano}{2006}]{lehmann2006testing}
Lehmann, E.~L. and J.~P. Romano (2006).
\newblock {\em Testing Statistical Hypotheses\/} (3rd ed.).
\newblock New York: Springer.

\bibitem[\protect\citeauthoryear{Little and Yau}{Little and
  Yau}{1998}]{little1998statistical}
Little, R.~J. and L.~H. Yau (1998).
\newblock {Statistical techniques for analyzing data from prevention trials:
  Treatment of no-shows using Rubin's causal model}.
\newblock {\em Psychological Methods\/}~{\em 3}, 147.

\bibitem[\protect\citeauthoryear{Liu}{Liu}{2004}]{liu2004robit}
Liu, C. (2004).
\newblock {\em Robit regression: a simple robust alternative to logistic and
  probit regression}, pp.\  227--238.
\newblock In {\it Applied Bayesian Modeling and Causal Inference From
  Incomplete-Data Perspectives} (A. Gelman and X. L. Meng, eds.), New York:
  Wiley.

\bibitem[\protect\citeauthoryear{Mattei and Mealli}{Mattei and
  Mealli}{2011}]{mattei2011augmented}
Mattei, A. and F.~Mealli (2011).
\newblock Augmented designs to assess principal strata direct effects.
\newblock {\em Journal of the Royal Statistical Society: Series B (Statistical
  Methodology)\/}~{\em 73}, 729--752.

\bibitem[\protect\citeauthoryear{Mealli and Pacini}{Mealli and
  Pacini}{2013}]{mealli2013using}
Mealli, F. and B.~Pacini (2013).
\newblock Using secondary outcomes to sharpen inference in randomized
  experiments with noncompliance.
\newblock {\em Journal of the American Statistical Association\/}~{\em 108},
  1120--1131.

\bibitem[\protect\citeauthoryear{Mealli, Pacini, and Stanghellini}{Mealli
  et~al.}{2016}]{mealli2016identification}
Mealli, F., B.~Pacini, and E.~Stanghellini (2016).
\newblock Identification of principal causal effects using additional outcomes
  in concentration graphs.
\newblock {\em Journal of Educational and Behavioral Statistics\/}~{\em 41},
  463--480.

\bibitem[\protect\citeauthoryear{Nelsen}{Nelsen}{2007}]{nelsen2007introduction}
Nelsen, R.~B. (2007).
\newblock {\em An Introduction to Copulas\/} (2nd ed.).
\newblock New York: Springer.

\bibitem[\protect\citeauthoryear{Pearl}{Pearl}{2000}]{judea2000causality}
Pearl, J. (2000).
\newblock {\em Causality: Models, Reasoning, and Inference\/} (2nd ed.).
\newblock Cambridge: Cambridge University Press.

\bibitem[\protect\citeauthoryear{Pearl}{Pearl}{2009}]{pearl2009letter}
Pearl, J. (2009).
\newblock Letter to the editor: Remarks on the method of propensity score.
\newblock {\em Statistics in Medicine\/}~{\em 28}, 1415--1416.

\bibitem[\protect\citeauthoryear{Qin, Gilbert, Follmann, and Li}{Qin
  et~al.}{2008}]{qin2008assessing}
Qin, L., P.~B. Gilbert, D.~Follmann, and D.~Li (2008).
\newblock Assessing surrogate endpoints in vaccine trials with case-cohort
  sampling and the cox model.
\newblock {\em The Annals of Applied Statistics\/}~{\em 2}, 386--407.

\bibitem[\protect\citeauthoryear{Rosenbaum}{Rosenbaum}{1984}]{rosenbaum1984consquences}
Rosenbaum, P.~R. (1984).
\newblock The consequences of adjustment for a concomitant variable that has
  been affected by the treatment.
\newblock {\em Journal of the Royal Statistical Society. Series A\/}~{\em 147},
  656--666.

\bibitem[\protect\citeauthoryear{Rosenbaum and Rubin}{Rosenbaum and
  Rubin}{1983}]{rosenbaum1983central}
Rosenbaum, P.~R. and D.~B. Rubin (1983).
\newblock The central role of the propensity score in observational studies for
  causal effects.
\newblock {\em Biometrika\/}~{\em 70}, 41--55.

\bibitem[\protect\citeauthoryear{Roy, Hogan, and Marcus}{Roy
  et~al.}{2008}]{roy2008principal}
Roy, J., J.~W. Hogan, and B.~H. Marcus (2008).
\newblock Principal stratification with predictors of compliance for randomized
  trials with 2 active treatments.
\newblock {\em Biostatistics\/}~{\em 9}, 277--289.

\bibitem[\protect\citeauthoryear{Rubin}{Rubin}{2004}]{rubin2004direct}
Rubin, D.~B. (2004).
\newblock Direct and indirect causal effects via potential outcomes (with
  discussion and reply).
\newblock {\em Scandinavian Journal of Statistics\/}~{\em 31}, 161--170.

\bibitem[\protect\citeauthoryear{Rubin}{Rubin}{2006}]{rubin2006causal}
Rubin, D.~B. (2006).
\newblock Causal inference through potential outcomes and principal
  stratification: application to studies with ``censoring" due to death (with
  discussion).
\newblock {\em Statistical Science\/}~{\em 21}, 299--309.

\bibitem[\protect\citeauthoryear{Schwartz, Li, and Mealli}{Schwartz
  et~al.}{2011}]{schwartz2011bayesian}
Schwartz, S.~L., F.~Li, and F.~Mealli (2011).
\newblock A bayesian semiparametric approach to intermediate variables in
  causal inference.
\newblock {\em Journal of the American Statistical Association\/}~{\em 106},
  1331--1344.

\bibitem[\protect\citeauthoryear{Sommer and Zeger}{Sommer and
  Zeger}{1991}]{sommer1991estimating}
Sommer, A. and S.~L. Zeger (1991).
\newblock On estimating efficacy from clinical trials.
\newblock {\em Statistics in Medicine\/}~{\em 10}, 45--52.

\bibitem[\protect\citeauthoryear{Stuart and Jo}{Stuart and
  Jo}{2015}]{stuart2015assessing}
Stuart, E.~A. and B.~Jo (2015).
\newblock Assessing the sensitivity of methods for estimating principal causal
  effects.
\newblock {\em Statistical Methods in Medical Research\/}~{\em 24}, 657--674.

\bibitem[\protect\citeauthoryear{Tingley, Yamamoto, Hirose, Keele, and
  Imai}{Tingley et~al.}{2014}]{tingley2014mediation}
Tingley, D., T.~Yamamoto, K.~Hirose, L.~Keele, and K.~Imai (2014).
\newblock Mediation: R package for causal mediation analysis.
\newblock {\em Journal of Statistical Software\/}~{\em 59}, 1--38.

\bibitem[\protect\citeauthoryear{VanderWeele}{VanderWeele}{2008}]{vanderweele2008simple}
VanderWeele, T.~J. (2008).
\newblock Simple relations between principal stratification and direct and
  indirect effects.
\newblock {\em Statistics and Probability Letters\/}~{\em 78}, 2957--2962.

\bibitem[\protect\citeauthoryear{Vinokur, Price, and Schul}{Vinokur
  et~al.}{1995}]{vinokur1995impact}
Vinokur, A.~D., R.~H. Price, and Y.~Schul (1995).
\newblock {Impact of the JOBS intervention on unemployed workers varying in
  risk for depression}.
\newblock {\em American Journal of Community Psychology\/}~{\em 23}, 39--74.

\bibitem[\protect\citeauthoryear{Vinokur and Schul}{Vinokur and
  Schul}{1997}]{vinokur1997mastery}
Vinokur, A.~D. and Y.~Schul (1997).
\newblock Mastery and inoculation against setbacks as active ingredients in the
  jobs intervention for the unemployed.
\newblock {\em Journal of Consulting and Clinical Psychology\/}~{\em 65}, 867.

\bibitem[\protect\citeauthoryear{Wang, Zhou, and Richardson}{Wang
  et~al.}{2017}]{wang2017identification}
Wang, L., X.-H. Zhou, and T.~S. Richardson (2017).
\newblock Identification and estimation of causal effects with outcomes
  truncated by death.
\newblock {\em Biometrika\/}~{\em 104}, 597--612.

\bibitem[\protect\citeauthoryear{Yang and Ding}{Yang and
  Ding}{2018}]{yang2018using}
Yang, F. and P.~Ding (2018).
\newblock Using survival information in truncation by death problems without
  the monotonicity assumption.
\newblock {\em Biometrics\/}~{\em 74}, 1232--1239.

\bibitem[\protect\citeauthoryear{Yuan, Feller, Miratrix, et~al.}{Yuan
  et~al.}{2019}]{yuan2018identifying}
Yuan, L.-H., A.~Feller, L.~W. Miratrix, et~al. (2019).
\newblock Identifying and estimating principal causal effects in a multi-site
  trial of early college high schools.
\newblock {\em The Annals of Applied Statistics\/}~{\em 13}, 1348--1369.

\bibitem[\protect\citeauthoryear{Zellner}{Zellner}{1976}]{zellner1976bayesian}
Zellner, A. (1976).
\newblock Bayesian and non-{B}ayesian analysis of the regression model with
  multivariate student-$t$ error terms.
\newblock {\em Journal of the American Statistical Association\/}~{\em 71},
  400--405.

\bibitem[\protect\citeauthoryear{Zhang and Rubin}{Zhang and
  Rubin}{2003}]{zhang2003estimation}
Zhang, J.~L. and D.~B. Rubin (2003).
\newblock Estimation of causal effects via principal stratification when some
  outcomes are truncated by ``death''.
\newblock {\em Journal of Educational and Behavioral Statistics\/}~{\em 28},
  353--368.

\bibitem[\protect\citeauthoryear{Zhang, Rubin, and Mealli}{Zhang
  et~al.}{2009}]{zhang2009likelihood}
Zhang, J.~L., D.~B. Rubin, and F.~Mealli (2009).
\newblock Likelihood-based analysis of causal effects of job-training programs
  using principal stratification.
\newblock {\em Journal of the American Statistical Association\/}~{\em 104},
  166--176.

\bibitem[\protect\citeauthoryear{Zigler and Belin}{Zigler and
  Belin}{2012}]{zigler2012bayesian}
Zigler, C.~M. and T.~R. Belin (2012).
\newblock A {B}ayesian approach to improved estimation of causal effect
  predictiveness for a principal surrogate endpoint.
\newblock {\em Biometrics\/}~{\em 68}, 922--932.

\end{thebibliography}

\newpage

\begin{center}
{\bf \huge 
Supplementary Materials}
\end{center}
\bigskip 

Appendix A provides the proofs of the theorems. 

Appendix B provides the proofs of the corollaries and propositions, and presents additional results for models related to $t$ distributions. Let
 $\bm{t}_p(\bm{\mu},\bm{\Sigma},\nu)$ denote the $p$-dimensional $t$ distribution with median $\bm\mu$, scale matrix $\bm\Sigma$, and degrees of freedom $\nu$, and let $T_{\nu}(\cdot)$ denote the cumulative distribution function of the standard $t$ distribution with degrees of freedom $\nu$.

Appendix C provides more details for the data analysis.

\section*{\bf Appendix A: Proofs of the theorems}
To prove the theorems, we need the following lemma from importance sampling.
\begin{lemma}
\label{lem::1}
 Let $f_X(x)$ and $f_Y(y)$ be the density functions of $X$ and Y. For any function $g(\cdot)$,  
\begin{eqnarray*}
\E \{g(X)\} = \E\left\{ \frac{f_X(Y)}{f_Y(Y)} g(Y)\right\},
\end{eqnarray*}
provided the existence of the moments.
\end{lemma}
 
\vspace{5mm}
\noindent{\it Proof of Theorem~\ref{thm::mon:pi}.} From the law of total expectation,  
\begin{eqnarray}
 \nonumber \E \left\{ \frac{e_{s_1}(W)}{e_{s_1}} \frac{ZY}{1-\pi(W)}\right\} &=& \E \left[ \E \left\{ \frac{e_{s_1}(W)}{e_{s_1}} \frac{ZY_1}{\pi(W)} \Bigg | W \right\} \right]\\
\nonumber &=& \E \left[  \frac{e_{s_1}(W)}{e_{s_1} \pi(W) } \E \left\{ ZY_1 \mid W \right\} \right]\\
 \nonumber &=& \E \left[  \frac{e_{s_1}(W)}{e_{s_1} \pi(W) } \E \left(Z\mid W \right) \E (Y_1\mid W ) \right]
\qquad (\text{Assumption~\ref{asm:randomization}})  
 \\
 \label{eqn::mon:pi1} &=& \E \left\{  \frac{e_{s_1}(W)}{e_{s_1}} \E (Y_1\mid W ) \right\}.
\end{eqnarray}
On the other hand, from the law of total expectation, 
\begin{eqnarray}
 \label{eqn::mon:pi2} \E (Y_1 \mid S_1=s_1) &=& \E \left\{ \E (Y_1 \mid W, S_1=s_1) \mid S_1=s_1 \right\}= \E \left\{ \E (Y_1 \mid W) \mid S_1=s_1 \right\},
\end{eqnarray}
where the last equality follows from Assumption~\ref{asm:pi}. The expectation in~\eqref{eqn::mon:pi1} is with respect to $W$ and the expectation in~\eqref{eqn::mon:pi2} is with respect to $W\mid S_1=s_1$. Therefore, from Lemma~\ref{lem::1},  
\begin{eqnarray*}
 \E \left\{ \E (Y_1 \mid W) \mid S_1=s_1 \right\} = \E \left\{  \frac{\pr(W\mid S_1=s_1)}{\pr(W)} \E (Y_0\mid W ) \right\}= \E \left\{  \frac{e_{s_1}(W)}{e_{s_1}} \E (Y_0\mid W ) \right\}.
\end{eqnarray*}
As a result, 
\begin{eqnarray*}
 \E \left\{ \frac{e_{s_1}(W)}{e_{s_1}} \frac{ZY}{\pi(W)}\right\} =\E (Y_1 \mid S_1=s_1) .
\end{eqnarray*}
Similarly, we can show
\begin{eqnarray*}
 \E \left\{ \frac{e_{s_1}(W)}{e_{s_1}} \frac{(1-Z)Y}{1-\pi(W)}\right\} =\E (Y_0 \mid S_1=s_1) .
\end{eqnarray*}
\QEDB

\vspace{5mm}
\noindent {\it Proof of Theorem~\ref{thm::mon:discrete}.} Because $\text{rank}(M^\top M)=K$, we can find an invertible $K \times K$ sub-matrix of $M$, denoted by $H$. Without loss of generality, we use $w_1,\ldots,w_K$ to denote the corresponding values of $W$ in $H$. First, for any $s_1$, $\pr(Y_1=y_1\mid S_1=s_1)=\pr(Y=y_1 \mid Z=1,S=s_1)$, which can be identified from the observed data. 
Assumption~\ref{asm:exclusion} implies
\begin{eqnarray}
\nonumber \pr(Y=y_0 \mid Z=0, W=w_k) 
 &=&\sum_{s_1} \pr(Y_0=y_0 \mid Z=0,S_1=s_1,W=w_j) \pr(S=s_1 \mid Z=0,W=w_k)\\
\label{eqn::thm1}&=&\sum_{s_1} \pr(Y_0=y_0 \mid S_1=s_1) \pr(S=s_1 \mid Z=0,W=w_k),
\end{eqnarray}
for $k=1,\ldots,K$, where $\pr(Y=y_0 \mid Z=0, W=w_k)$ and $\pr(S=s_1 \mid Z=0,W=w_k)$ can be identified from the observed data. Because $H$ is invertible, we can solve \eqref{eqn::thm1} to obtain $\pr(Y_0=y_0 \mid S_1=s_1)$ for all $s_1$. Therefore, the PCEs are identifiable. \QEDB

\vspace{5mm}
\noindent {\it Proof of Theorem~\ref{thm::constant:complete}.} From Assumptions~\ref{asm:randomization}~and~\ref{asm:exclusion}, we have 
\begin{eqnarray}
\label{eqn::thm3-Y0}
 \pr(Y=y_0 \mid Z=0, W)=\pr(Y_0=y_0 \mid W) = \E\{\pr(Y_0=y_0\mid S_1)\mid W\}
\end{eqnarray}
for any fixed $y_0$. Since $ \pr(Y=y_0 \mid Z=0, W)$ is identifiable from the observed data, it suffices to show that $\pr(Y_0=y_0\mid S_1)$ is uniquely determined by~\eqref{eqn::thm3-Y0}.
If there exist two functions of $S_1$, $f_1(Y_0=y_0\mid S_1)$ and $f_2(Y_0=y_0\mid S_1)$, satisfying
\begin{eqnarray*}
&&\E\{f_{1}(Y_0=y_0\mid S_1)\mid W\}=\E\{f_{2}(Y_0=y_0\mid S_1)\mid W\},
 \end{eqnarray*}
 then $\E\{g(S_1)\mid W\}=0$, where $g(S_1)=f_{1}(Y_0=y_0\mid S_1)-f_{2}(Y_0=y_0\mid S_1)$. From the definition of completeness, $g(S_1)=0$ and hence the distribution of $\pr(Y_0=y_0\mid S_1)$ is identifiable. Therefore, the PCEs are identifiable. \QEDB
  
\vspace{5mm}
\noindent {\it Proof of Theorem~\ref{cor::constant:exp}.}  From the completeness of exponential family (shown in Result \ref{re:ex} in the next section), $\mathcal{P}_{\mathcal{W}}$ is complete. Then from Theorem~\ref{thm::constant:complete}, the PCEs are identifiable.  \QEDB
\vspace{5mm}

\noindent {\it Proof of Theorem~\ref{thm:semipara}.}  From Lemma~\ref{lem::semipara:com}, $\mathcal{P}_{\mathcal{W}}$ is complete. Then from Theorem~\ref{thm::constant:complete}, the PCEs are identifiable.  \QEDB
\vspace{5mm}

\noindent{\it Proof of Theorem~\ref{thm::pi}.} Similar to the proof of Theorem~\ref{thm::mon:pi}, we can show that 
\begin{eqnarray*}
\E \left\{ \frac{e_{s_1,s_0}(W)}{e_{s_1,s_0}}\cdot \frac{ZY}{\pi(W)}\right\}&=& \E(Y_1 \mid S_1=s_1,S_0=s_0),\\
 \E \left\{ \frac{e_{s_1,s_0}(W)}{e_{s_1,s_0}} \cdot \frac{(1-Z)Y}{1-\pi(W)}\right\} &=& \E(Y_0 \mid S_1=s_1,S_0=s_0).
\end{eqnarray*}
 \QEDB

\vspace{5mm}

\noindent {\it Proof of Theorem~\ref{thm::nonmon:discrete}.} Because $\text{rank}(M_{s_0}^\top M_{s_0})=K$, we can find an invertible $K \times K$ sub-matrix of $M_{s_0}$, denoted by $H_{s_0}$, with the corresponding values of $W$ denoted by $w_1,\ldots,w_K$. For any fixed $s_0$, 
Assumption~\ref{asm:exclusion} implies
\begin{eqnarray}
\nonumber&&\pr(Y=y_0 \mid Z=0, S=s_0,W=w_j) \\
\nonumber&=&\sum_{s_1} \pr(Y_0=y_0 \mid Z=0,S_1=s_1,S_0=s_0,W=w_j) \pr(S_1=s_1 \mid S_0=s_0,W=w_j)\\
\label{eqn:thm5}&=&\sum_{s_1} \pr(Y_0=y_0 \mid S_1=s_1,S_0=s_0) \pr(S_1=s_1 \mid S_0=s_0,W=w_j)
\end{eqnarray}
for $j=1,\ldots,K$,
where $\pr(Y=y_0 \mid Z=0, S=s_0,W=w_j)$ and $\pr(S_1=s_1 \mid S_0=s_0,W=w_j)$ can be identified from the observed data. Because $H_{s_0}$ is invertible, we can solve \eqref{eqn:thm5} to obtain $\pr(Y_0=y_0 \mid S_1=s_1,S_0=s_0)$ for all $s_1$. Similarly, we can identify $\pr(Y_1=y_1 \mid S_1=s_1,S_0=s_0)$ for all $s_0$ under $\text{rank}(M_{s_1}^\top M_{s_1})=K$. Therefore, the PCEs are identifiable under Conditions (a) and (b).  \QEDB

\vspace{5mm}

\noindent {\it Proof of Theorem~\ref{thm:noncon:complete}.}  From Assumptions~\ref{asm:randomization}~and~\ref{asm:exclusion}, we have 
\begin{eqnarray}
\label{eqn::thm8-Y0}
\nonumber \pr(Y=y_0 \mid Z=0,S=s_0, W)&=&\pr(Y_0=y_0 \mid S_0=s_0,W) \\
 &=& \E\{\pr(Y_0=y_0\mid S_1,S_0=s_0)\mid S_0=s_0,W\}
\end{eqnarray}
for any fixed $y_0$ and $s_0$. Since $ \pr(Y=y_0 \mid Z=0, S=s_0,W)$ is identifiable from the observed data, it suffices to show that $\pr(Y_0=y_0\mid S_1)$ is uniquely determined by~\eqref{eqn::thm3-Y0}.
If there exist two functions of $S_1$, $f_{1}(Y_0=y_0\mid S_1, S_0=s_0)$ and $f_{2}(Y_0=y_0\mid S_1, S_0=s_0)$, satisfying
\begin{eqnarray*}
&& \E\{f_{1}(Y_0=y_0\mid S_1, S_0=s_0)\mid W\}=\E\{f_{2}(Y_0=y_0\mid S_1, S_0=s_0)\mid W\},
\end{eqnarray*}
 then  $\E\{g_{s_0}(S_1)\mid S_0=s_0, W\}=0$, where $g_{s_0}(S_1)=f_{1}(Y_0=y_0\mid S_1, S_0=s_0)-f_{2}(Y_0=y_0\mid S_1, S_0=s_0)$. Because $\pr(S_1\mid S_0=s_0,W)$ is identifiable and $\mathcal{P}_{\mathcal{W},s_0}$ is complete, $g_{s_0}(S_1)=0$. Therefore, the distribution of $\pr(Y_0=y_0\mid S_1, S_0=s_0)$ is identifiable. Similarly, the distribution of $\pr(Y_1=y_1\mid S_1=s_1, S_0)$ is identifiable for any fixed $y_1$ and $s_1$ if $\mathcal{P}_{\mathcal{W},s_1}$ is complete. Therefore, the PCEs are identifiable under Conditions (a) and (b). \QEDB

\section*{\bf Appendix B: Proof of the corollaries and propositions}
\subsection*{\bf Appendix B.1: Proofs of the completeness of some distribution families}
\setcounter{result}{0}
\renewcommand {\theresult} {S\arabic{result}}

In this section, we show the completeness of exponential family and three distribution families based on Normal, $t$, and Logistic distribution, respectively.

\setcounter{lemma}{0}
\renewcommand {\thelemma} {A.\arabic{lemma}}
\begin{result}\citep[][Theorem~4.3.1]{lehmann2006testing}
\label{re:ex}
Let $X$ be a random vector with probability distribution
\begin{eqnarray*}
\text{d} \mathbb{P}_{\theta}(x)=C(\theta)\exp \left\{ \sum_{j=1}^J\theta_jt_j(x)\right\} \text{d} \mu(x),
\end{eqnarray*}
and let $\mathcal{P}_\omega$ be the family of distributions of $t=(t_1(X),\ldots,t_J(X))$ as $\bm{\theta}=(\theta_1,\ldots,\theta_J)$ varies over the set $\omega$. Then $\mathcal{P}_\omega$ is complete provided $\omega$ contains a $J$-dimensional rectangle in $\mathbb{R}^J$.
\end{result}


\begin{result}
\label{re:app1}
Suppose that $S_1\overset{\text{d}}{=}h(W)+\sigma(W)\epsilon$ with $\epsilon \ind W$, and $h(w)$ and $\sigma(w)$ continuously differentiable. If $\epsilon \sim N(0,1)$, then $\mathcal{P}_{\mathcal{W}}$ is complete.
\end{result}
 \noindent {\it Proof of Result \ref{re:app1}.} We verify the conditions in Lemma \ref{lem::semipara:com}. First, because $\phi(t)=\exp(-t^2/2)$, Condition (a) holds. Second, from the property of the Normal distribution, it is easy to show that Condition (b) holds. Finally, based on the identifiability of Normal mixture models \citep{everitt1981finite}, Condition (c) holds. Thus, from Lemma \ref{lem::semipara:com}, $\mathcal{P}_{\mathcal{W}}$ is complete.\QEDB
 
 \vspace{5mm}
 
 \begin{result}
 \label{re:app2}
Suppose $S_1\overset{\text{d}}{=}h(W)+\sigma(W)\epsilon$ with $\epsilon \ind W$, and $h(w)$ and $\sigma(w)$ continuously differentiable. If $\epsilon \sim \bm{t}_1(0,1,\nu)$, then $\mathcal{P}_{\mathcal{W}}$ is complete.
\end{result}
 \noindent {\it Proof of Result \ref{re:app2}.} We verify the conditions in Lemma \ref{lem::semipara:com}. First, the characteristic function of $\epsilon$ is 
 \begin{eqnarray*}
\phi(t)=\frac{K_{\nu/2}(\sqrt{\nu}|t|)\cdot (\sqrt{\nu}|t|)^{\nu/2}}{\Gamma(\nu/2)2^{\nu/2-1}},
\end{eqnarray*}
where $\Gamma(\cdot)$ is the Gamma function and $K_{\nu/2}(\cdot)$ is the modified Bessel function of the second kind. \citet{abramowitz1966handbook} ensures that
\begin{eqnarray*}
\lim_{t \rightarrow \infty} \frac{K_{\nu/2}(\sqrt{\nu}|t|)}{\exp(-\sqrt{\nu}|t|)/\sqrt{2\sqrt{\nu}|t|}}=1.
\end{eqnarray*}
Therefore, the dominating term in $\phi(t)$ is $\exp(-\sqrt{\nu}|t|)$, and hence Condition (a) holds. Second, 
 it is easy to verify Condition (b). Finally, from the identifiability of $t$ mixture model \citep{everitt1981finite}, Condition (c) holds. Thus, from Lemma \ref{lem::semipara:com}, $\mathcal{P}_{\mathcal{W}}$ is complete.\QEDB
 
 \vspace{5mm}
 
 \begin{result}
 \label{re:app3}
Suppose $S_1\overset{\text{d}}{=}h(W)+\sigma(W)\epsilon$ with $\epsilon \ind W$, and $h(w)$ and $\sigma(w)$ continuously differentiable. If $\epsilon$ follows a standard Logistic distribution with density
$
 {\exp(-\epsilon)}/ {\{1+\exp(-\epsilon)\}^2},
$
 then $\mathcal{P}_{\mathcal{W}}$ is complete.
\end{result}
 \noindent {\it Proof of Result \ref{re:app3}.} We verify the conditions in Lemma \ref{lem::semipara:com}. First, the characteristic function of $\epsilon$ is 
$
\phi(t)= {\pi t} / {\sinh(\pi t)}. 
$
Because the dominating term in $\phi(t)$ is $\exp(-\pi |t|)$, Condition (a) holds. Second, it is easy to show that Condition (b) holds. Finally, from the identifiability of Logistic mixture models \citep{everitt1981finite}, Condition (c) holds. Thus, from Lemma \ref{lem::semipara:com}, $\mathcal{P}_{\mathcal{W}}$ is complete. \QEDB

\subsection*{\bf Appendix B.2: Proof of the corollaries}

\noindent {\it Proof of Corollary~\ref{prop::bivariateN}.}  
Suppose that the distribution of $(S_1, W)$ is bivariate Normal with means $(\mu_S, \mu_W)$, variances $(\sigma_S^2, \sigma_W^2)$, and correlation coefficient $\rho$. From the bivariate Normal distribution of $(S_1,W)$, we have
\begin{eqnarray*}
\pr(S_1=s_1 \mid W=w) &=& \pr(S_1=s_1 \mid W=w) \\
&=&\frac{1}{\sqrt{2 \pi \sigma^2(w)}}\exp \left[-\frac{\{s_1-\mu(w)\}^2}{2\sigma^2(w)}\right]\\
&=& \frac{1}{\sqrt{2 \pi \sigma^2(w)}} \exp \left\{- \frac{\mu^2(w)}{2\sigma^2(w)} \right\} \exp \left\{ -\frac{s_1^2}{2\sigma^2(w)} \right\} \exp \left\{ \frac{s_1 \mu(w)}{\sigma^2(w)} \right\},
\end{eqnarray*}
where $\mu(w) =\mu_S+\rho \sigma_S /\sigma_W(w-\mu_W)$ and $\sigma^2(w)=\sigma_S^2(1-\rho^2)$. From Theorem~\ref{cor::constant:exp}, we have $t(s_1)=s_1$ and $\eta(w)=\mu(w)/\sigma^2(w)$, and thus the PCEs are identifiable. \QEDB
\vspace{5mm}

\noindent {\it Proof of Corollary~\ref{cor:trivariateN}.} First, we can identify $\{ \mu_1(w),\mu_0(w),\sigma_1^2(w),\sigma_0^2(w) \} $ from the observed data. From the bivariate Normal assumption,
the conditional distribution of $S_1 \mid (S_0=s_0,W=w)$ is a Normal distribution with mean $\mu(s_0,w)=\mu_1(w)+\rho(w) \sigma_1(w)/\sigma_0(w)\{s_0-\mu_0(w)\}$ and variance $\sigma^2(w)=\{1-\rho^2(w)\}\sigma^2_1(w)$, i.e., 
\begin{eqnarray*}
S_1\overset{\text{d}}{=}\mu(s_0,W)+\sigma(W) \epsilon, 
\end{eqnarray*}
where $\epsilon \ind W$ and $\epsilon \sim N(0,1)$. From Result \ref{re:app1}, $\mathcal{P}_{\mathcal{W},s_0}$ is complete. Similarly,  $\mathcal{P}_{\mathcal{W},s_1}$ is also complete. Therefore, from Theorem~\ref{thm:noncon:complete}, the PCEs are identifiable. \QEDB
\vspace{5mm}

\subsection*{\bf Appendix B.3: Proof of the propositions}

\noindent {\it Proof of Proposition~\ref{ex::nind:linear}.} First, because $S_1$ is observed when $Z=1$, we can identify $\pr(Y_1\mid S_1)$ and $g(W)=\E(S_1\mid W)$. Second, the linear models of $S_1$ and $Y_0$ imply 
\begin{eqnarray*}
Y_0=\beta_0+\alpha g(W)+\sum_{j=1}^J \beta_j f_j(W)+\sigma_1(W)\epsilon_1+\sigma_2(W)\epsilon_2.
\end{eqnarray*}
 From the classic result of linear models, the parameters are identifiable
because $(\epsilon_1, \epsilon_2) \ind W$ and $\{1, g(w), f_1(w),\ldots,f_J(w)\}$ are linearly independent. Therefore, we can identify $\pr(Y_0\mid S_1)$ and hence the PCEs. \QEDB
\vspace{5mm}

\noindent {\it Proof of Proposition~\ref{ex::nind:probit}.} First, because we can observed $S_1$ when $Z=1$, we can identify $\pr(Y_1\mid S_1)$. Second, the distribution of $(S_1,W)$ is identifiable, so are $g(w)$ and $\sigma^2$. Third, by the law of total probability,
\begin{eqnarray*}
\pr(Y=1\mid Z=0,W=w)
&=&\int \pr(Y_0=1\mid Z=0,S_1=s,W=w)\pr(S_1=s \mid W=w)\text{d}s\\
&=&\int \pr(Y_0=1\mid S_1=s,W=w)\pr(S_1=s \mid W=w)\text{d}s\\
&=&\int \Phi\left\{\alpha s+\sum_{j=1}^J \beta_j f_j(W)\right\}\pr(S_1=s \mid W=w)\text{d}s\\
&=&\Phi\left\{\frac{\beta_0+\alpha g(w)+\sum_{j=1}^J \beta_j f_j(w)}{\sqrt{1+\alpha^2\sigma^2}}\right\}.
\end{eqnarray*}
Because $\{1,g(w), f_1(w),\ldots,f_J(w)\}$ are linearly independent, from the identifiability of the Probit model, we can identify 
\begin{eqnarray*}
\frac{\alpha}{\sqrt{1+\alpha^2\sigma^2}}, \frac{\beta_0}{\sqrt{1+\alpha^2\sigma^2}}, \ldots, \frac{\beta_J}{\sqrt{1+\alpha^2\sigma^2}}.
\end{eqnarray*}
Since $\sigma^2$ is identifiable, we can identify $\{ \alpha,\beta_0,\ldots,\beta_J \} $ and hence the PCEs.
 \QEDB
\vspace{5mm}

\noindent {\it Proof of Proposition~\ref{ex:noncon:noind:binary}.} First, we can identify the joint distribution of $(S_1,S_0 )$ given $W$:
\begin{eqnarray*}
&&\pr(S_1=1,S_0=1\mid W=w) = \pr(S_1=1,S_0=1\mid Z=0, W=w)= \pr(S=1\mid Z=0, W=w),\\
&&\pr(S_1=0,S_0=0\mid W=w) = \pr(S_1=0,S_0=0\mid Z=1, W=w)= \pr(S=0\mid Z=1, W=w),\\
&&\pr(S_1=1,S_0=0\mid W=w) =1-\pr(S_1=1,S_0=1\mid W=w)-\pr(S_1=0,S_0=0\mid W=w).
\end{eqnarray*}
Then,  
\begin{eqnarray*}
 \pr(S_0=1\mid S_1=1,W=w) = \frac{\pr(S=1\mid Z=0,W=w)}{\pr(S=1\mid Z=1,W=w)},\quad 
 \pr(S_0=1\mid S_1=0,W=w) =0.
\end{eqnarray*}
Because the subpopulation with $(Z=1,S=1)$ is a mixture of the subpopulations with $(S_1=1,S_0=1)$ and $(S_1=1,S_0=0)$, we have
\begin{eqnarray}
\nonumber \E(Y \mid Z=1,S=1,W=w) 
&=& \E(Y_1 \mid S_1=1,S_0=1,W=w)\pr(S_0=1\mid S_1=1,W=w)\\
\nonumber && +\E(Y_1 \mid S_1=1,S_0=0,W=w)\pr(S_0=0\mid S_1=1,W=w)\\
\label{eq:1} &=&
 \beta_{10}+\beta_{11}+ \beta_{12}\frac{\pr(S=1\mid Z=0,W=w)}{\pr(S=1\mid Z=1,W=w)}+\beta_{13}w.
\end{eqnarray}
The subpopulation with $(Z=1,S=0)$ is the same as the subpopulation with $(S_1=0,S_0=0)$, and thus 
\begin{eqnarray}
\label{eq:2} &&\E(Y \mid S=0,Z=1,W=w) = \E(Y_1 \mid S_1=0,S_0=0,W=w)=\beta_{10}+\beta_{13}w.
\end{eqnarray}
Because $\pr(S=1\mid Z=0,W=w)/\pr(S=1\mid Z=1,W=w)$ is not constant in $w$, we can identify $(\beta_{10},\beta_{11},\beta_{12},\beta_{13})$ from \eqref{eq:1} and \eqref{eq:2}. Similarly, we can identify $(\beta_{00},\beta_{01},\beta_{02},\beta_{03})$ and hence the PCEs. \QEDB

\vspace{5mm}

\noindent {\it Proof of Proposition~\ref{ex:noncon:linear}.} From the linear model for $Y_1$,
\begin{eqnarray*}
&&\E(Y \mid Z=1,S=s_1,W=w)\\
&=& \int \E(Y_1 \mid S_1=s_1,S_0=s,W=w)\pr(S_0=s_0\mid S_1=s_1,W=w)\text{d}s\\
&=& \int\{ \beta_0+\alpha_1 s_1+\alpha_0 s_0+\sum_{j=1}^J \beta_j f_j(W)\}\pr(S_0=s_0\mid S_1=s_1,W=w)\text{d}s\\
&=&\beta_0+\alpha_1 s_1+\alpha_0 \E(S_0\mid S_1=s_1,W=w)+\sum_{j=1}^J \beta_j f_j(w).
\end{eqnarray*}
Because $\{1,s_1,\E(S_0\mid S_1=s_1,W=w), f_1(w),\ldots,f_J(w)\}$ are linearly independent, we can identify $(\beta_0,\ldots,\beta_J,\alpha_1,\alpha_0)$. Similarly, we can identify $(\beta'_0,\ldots,\beta'_J,\alpha'_1,\alpha'_0)$ and hence the PCEs. \QEDB

\vspace{5mm}

\noindent {\it Proof of Proposition~\ref{ex::noncon:nind:probit}.} We can first identify $\{ \mu_1(w),\mu_0(w),\sigma_1^2(w),\sigma_0^2(w) \} $. From the bivariate Normality, the conditional distribution $S_0 \mid (S_1=s_1,W=w)$ is Normal with mean $\mu(s_1,w)=\mu_0(w)+\rho(w) \sigma_0(w)/\sigma_1(w)\{s_1-\mu_1(w)\}$ and variance $\sigma^2(w)=\{1-\rho^2(w)\}\sigma^2_0(w)$.
Then, by the law of total probability, 
\begin{eqnarray*}
&&\pr(Y=1\mid Z=1,S=s_1,W=w)\\
&=&\int \pr(Y_1=1\mid S_1=s_1,S_0=s_0,W=w)\pr(S_0=s_0 \mid S_1=s_1, W=w)\text{d}s_0\\
&=&\int \Phi\left\{\beta_0+\alpha_1 s_1+\alpha_0 s_0+\sum_{j=1}^J \beta_j f_j(W)\right\}\pr(S_0=s_0 \mid S_1=s_1, W=w)\text{d}s_1\\
&=&\Phi\left\{\frac{\beta_0+\alpha_1s_1 +\alpha_0 \mu(s_1,w)+\sum_{j=1}^J \beta_j f_j(w)}{\sqrt{1+\alpha_0^2\sigma^2(w)}}\right\}.
\end{eqnarray*}
Because $\{1,s_1, \E(S_0\mid S_1=s_1,W=w), f_1(w),\ldots,f_J(w)\}$ are linearly independent, we can use Probit regression to first identify $\alpha_1$ and $\alpha_0$ by fixing $w$ and then identify $\beta_j$ for $j=1,\ldots,J$ by fixing $s_1$. Similarly, we can identify $(\beta'_0,\ldots,\beta'_J,\alpha'_1,\alpha'_0)$ and hence the PCEs. \QEDB
\vspace{5mm}

\subsection*{\bf Appendix B.4: Models based on $t$ distributions}
\setcounter{corollary}{0}
\renewcommand {\thecorollary} {S\arabic{corollary}}
We give identification results for models based on $t$ distributions.
 These models are frequently applied for robust analysis when the data have heavier tails than the standard Normal distribution \citep{zellner1976bayesian,lange1989robust, liu2004robit, gelman2014bayesian}.  

We first give the results under Assumption~\ref{asm:exclusion}.
The following result is an application of Theorem~\ref{thm:semipara} to models based on $t$ distributions with constant control intermediate variable.
\begin{corollary}
\label{cor:bivariateT}
Suppose that $W$ is continuous and 
Assumptions~\ref{asm:randomization},~\ref{asm:exclusion},~and~\ref{asm:constant} hold, 
and  $S_1 \mid W=w \sim t_1( h(w), \sigma(w),\nu)$ with unknown $\nu$, where  $h(w)$ and $\sigma(w)$ are continuously differentiable functions that can be unknown. The PCEs are identifiable. 
\end{corollary}
\noindent {\it Proof of Corollary~\ref{cor:bivariateT}.}  
From Result \ref{re:app2}, $\mathcal{P}_{\mathcal{W}}$ is complete. Thus, from Theorem~\ref{thm:semipara}, the PCEs are identifiable. \QEDB
\vspace{5mm}

Similarly,  the following result is an application of Theorem~\ref{thm:noncon:complete} to models based on $t$ distributions with non-constant control intermediate variable.
\begin{corollary}
\label{cor:trivariateT}
For a continuous $W$,
suppose that Assumptions \ref{asm:randomization} and \ref{asm:exclusion} hold. If  
\begin{eqnarray*}
(S_1,S_0)\mid W=w \sim \bm{t}_2 \left\{ 
\begin{pmatrix}
\mu_1(w)\\
\mu_0(w)
\end{pmatrix}, \begin{pmatrix}
 \sigma^2_{1}(w)& \rho(w)\sigma_{1}(w)\sigma_{0}(w)\\
\rho(w)\sigma_{1}(w)\sigma_{0}(w) & \sigma^2_0(w)
\end{pmatrix},  \nu
 \right\},
\end{eqnarray*}
with known values of $\rho(w)$, then the PCEs are identifiable.
\end{corollary}
\noindent {\it Proof of Corollary~\ref{cor:trivariateT}.}  
First, we can identify $\{ \mu_1(w),\mu_0(w),\sigma_1^2(w),\sigma_0^2(w) \} $. For any fixed $s_0$,  \citet[][Conclusion One]{ding2016conditional} implies that $S_1\overset{\text{d}}{=}h_{s_0}(W)+\sigma_{s_0}(W)\epsilon$, where
\begin{eqnarray*}
&&h_{s_0}(W)=\mu_1(W)+\frac{\sigma_{10}(W)}{\sigma^2_0(W)}\{s_0-\mu_0(W)\},\\
&&\sigma_{s_0}(W)=\sqrt{\frac{\nu+\{s_0-\mu_0(W)\}/\sigma^2_0(W)}{\nu+1}} \cdot \sqrt{\sigma_0(W)-\frac{\sigma_{10}(W)}{\sigma^2_0(W)}},\\
&&\epsilon \sim t_1(0,1,\nu+1), \quad \epsilon \ind W.
\end{eqnarray*}
From Result \ref{re:app2}, $\mathcal{P}_{\mathcal{W},s_0}$ is complete. Similarly, $\mathcal{P}_{\mathcal{W},s_1}$ is also complete. Therefore, from Theorem~\ref{thm:noncon:complete}, the PCEs are identifiable.\QEDB

\vspace{5mm}

We then give the results when neither Assumption~\ref{asm:pi}~or~\ref{asm:exclusion} holds.
The following proposition extends the Probit model in Proposition~\ref{ex::nind:probit} to the Robit model \citep{liu2004robit} with a constant control intermediate variable.

\begin{proposition}
\label{ex:nind:robit}
Under Assumptions~\ref{asm:randomization}~and~\ref{asm:constant},
assume 
\begin{eqnarray*}
&& S_1= g(W)+\epsilon_{S_1},\\
&& Y_0=I\left\{ \beta_0+\alpha S_1+\sum_{j=1}^J \beta_j f_j(W)+\epsilon_{Y_0}>0 \right\},\\
&&(\epsilon_{S_1},\epsilon_{Y_0}) \sim  \bm{t}_2(\bm{\mu},\bm{\Sigma},\nu),
\end{eqnarray*}
where  $(\epsilon_{S_1},\epsilon_{Y_0}) \ind W$, $\bm{\mu}=(0,0)^\top$, $\bm{\Sigma}=\text{diag}(\sigma^2,1)$, and $g(w)$ can be unknown. If $\{1,g(w), f_1(w),\ldots,f_J(w)\}$ are linearly independent, then the PCEs are identifiable.
\end{proposition}

Proposition~\ref{ex::nind:probit} does not assume a joint model for $(S_1,Y_0)$. In contrast, Proposition \ref{ex:nind:robit} assumes a joint model for these two variables by restricting the error terms to follow a bivariate $t$ distribution. This is because two independent Normal random variables make a bivariate Normal random variable, but two independent $t$ random variables do not make a bivariate $t$ random variable even with the same degrees of freedom. 

\medskip

\noindent {\it Proof of Proposition~\ref{ex:nind:robit}.}
First, because $S_1$ is observed when $Z=1$, we can identify $\pr(Y_1\mid S_1)$. Second, because the distribution of $(S_1,W)$ is identifiable, we can identify $g(w)$ and $\sigma^2$. Then, by the law of total probability, 
\begin{eqnarray*}
 \pr(Y=1\mid Z=0,W=w) 
&=&\int \pr(Y=1\mid Z=0,S_1=s,W=w) \pr(S_1=s\mid Z=0) \text{d}s\\
&=& \int \pr(Y_0=1\mid S_1=s,W=w)\pr(S_1=s\mid Z=0) \text{d} s\\
 &=& T_{\nu} \left \{ \frac{\beta_0+\alpha g(W) +\sum_{j=1}^J \beta_j f_j(W)}{\sqrt{1+\alpha^2\sigma^2}}\right\}.
\end{eqnarray*}
Because $\{1,g(w), f_1(w),\ldots,f_J(w)\}$ are linearly independent, from the identifiability of the Robit model, we can identify 
\begin{eqnarray*}
\frac{\alpha}{\sqrt{1+\alpha^2\sigma^2}}, \frac{\beta_0}{\sqrt{1+\alpha^2\sigma^2}}, \ldots, \frac{\beta_J}{\sqrt{1+\alpha^2\sigma^2}}.
\end{eqnarray*}
Since $\sigma^2$ is identifiable, we can identify $\{ \alpha,\beta_0,\ldots,\beta_J \} $ and hence the PCEs.
 \QEDB

\vspace{5mm}

Similarly, for the case with a non-constant control intermediate variable,
we give an identification result for Robit outcome models as an extension of the Probit outcome models.

 \begin{proposition}
 \label{ex:noncon:nind:robit}
Under Assumption \ref{asm:randomization}, assume that  
\begin{eqnarray*}
&& S_1= g_1(W)+\epsilon_{S_1},
\quad S_0= g_0(W)+\epsilon_{S_0},\\
&& Y(1)=I\left\{ \beta_0+\alpha_1 S_1+\alpha_0 S_0+\sum_{j=1}^J \beta_j f_j(W)+\epsilon_{Y_1}>0\right\},\\
&& Y(0)=I\left\{\beta'_0+\alpha'_1 S_1+\alpha'_0 S_0+\sum_{i=1}^k \beta'_i h_i(W)+\epsilon_{Y_0}>0\right\},
\end{eqnarray*}
where $(\epsilon_{S_1},\epsilon_{S_0},\epsilon_{Y_1},\epsilon_{Y_0}) \sim  \bm{t}_4(\bm{\mu},\bm{\Sigma},\nu)$, $(\epsilon_{S_1},\epsilon_{S_0},\epsilon_{Y_1},\epsilon_{Y_0}) \ind W$, and
\begin{eqnarray*}
\bm{\mu}=\begin{pmatrix}0\\0\\0\\0 \end{pmatrix}, \quad \bm{\Sigma} = \begin{pmatrix}\sigma_{S_1}^2 & \sigma_{S_1S_0}&\sigma_{S_1Y_1}&0\\\sigma_{S_0S_1} & \sigma^2_{S_0}&0&\sigma_{S_0Y_0} \\\sigma_{Y_1S_1} & 0&\sigma^2_{Y_1}&\sigma_{Y_1Y_0}\\0& \sigma_{Y_0S_0}&\sigma_{Y_0Y_1}&\sigma^2_{Y_0} \end{pmatrix} , 
\end{eqnarray*}
with known $\sigma_{S_1S_0}$. The PCEs are identifiable, if $\{1,g_1(w), f_1(w),\ldots,f_J(w)\}$ are linearly independent, and $\{1, g_0(w)$, $h_1(w),\ldots,h_J(w)\}$ are linearly independent. 
\end{proposition}
\noindent {\it Proof of Proposition~\ref{ex:noncon:nind:robit}.} We can first identify $g_1(\cdot)$, $g_0(\cdot)$ and all the terms in $\bm{\Sigma}$ except $\sigma_{Y_0Y_1}$.
Then 
\begin{eqnarray*}
&&\pr(Y=1 \mid Z=1,S=s_1,W=w)\\
&=&\pr(\beta_0+\alpha_1 S_1+\alpha_0 S_0+\sum_{j=1}^J \beta_j f_j(W)+\epsilon_{Y_1}\geq 0 \mid S_1=s_1,W=w)\\
&=&\pr \left\{\beta_0+\alpha_1 s_1+\alpha_0 g_0(W)+\sum_{j=1}^J \beta_j f_j(W)+\epsilon_{Y_1}+\alpha_0\epsilon_{S_0}\geq 0\Bigg |  g_1(W)+\epsilon_{S_1}=s_1,W=w\right\}\\
&=&\pr \left[\epsilon_{Y_1}+\alpha_0\epsilon_{S_0} \geq -\left\{\beta_0+\alpha_1 s_1+\alpha_0 g_0(W)+\sum_{j=1}^J \beta_j f_j(W) \right\} \Bigg | \epsilon_{S_1}=s_1-g_1(W),W=w\right].
\end{eqnarray*}
\citet[][Conclusion One]{ding2016conditional} implies 
\begin{eqnarray*}
(\epsilon_{Y_1},\epsilon_{S_0})\mid \epsilon_{S_1}=x \sim \bm{t}_2\left(\bm{\mu}_1, \frac{\nu+d_1}{\nu+1}\bm{\Sigma}_1,\nu+1\right),
\end{eqnarray*}
 where
\begin{eqnarray*}
&&\bm{\mu}_1 =\begin{pmatrix}\sigma_{Y_1S_1}x/\sigma_{S_1}^2 \\ \sigma_{S_0S_1}x/\sigma_{S_1}^2 \end{pmatrix}, \quad \bm{\Sigma}_1=\begin{pmatrix}\sigma_{Y_1}^2-\sigma_{Y_1S_1}^2/\sigma_{S_1}^2 & -\sigma_{Y_1S_1}\sigma_{S_1S_0}/\sigma_{S_1}^2\\
-\sigma_{Y_1S_1}\sigma_{S_1S_0}/\sigma_{S_1}^2 & \sigma_{S_0}^2-\sigma_{S_1S_0}^2/\sigma_{S_1}^2 \end{pmatrix},\quad d_1=x^2/\sigma_{S_1}^2.
\end{eqnarray*}
Thus, 
\begin{eqnarray*}
\epsilon_{Y_1}+\alpha_0\epsilon_{S_0} \mid \epsilon_{S_1}=x \sim t_1\left(\alpha_0\frac{\sigma_{S_0S_1}}{\sigma^2_{S_1} }+\frac{\sigma_{Y_1S_1}}{\sigma^2_{S_1} }, \frac{\nu+d_1}{\nu+1}r(\alpha_0),\nu+1\right),
\end{eqnarray*}
where $r(\alpha_0) =(1,\alpha_0)\bm{\Sigma}_1(1,\alpha_0)^\top$. Therefore, 
\begin{eqnarray*}
&&\pr(Y=1 \mid Z=1,S=s_1,W=w)\\
&=& T_{\nu+1} \left[\frac{\beta_0+\alpha_1s_1+\alpha_0 g_0(w)-\left (\alpha_0\frac{\sigma_{S_0S_1}}{\sigma^2_{S_1} }+\frac{\sigma_{Y_1S_1}}{\sigma^2_{S_1} }\right) \{s_1-g_1(w)\}+\sum_{j=1}^J \beta_j f_j(W)}{\sqrt{t(s_1,w) r(\alpha_0)}}\right],
\end{eqnarray*}
where  
\begin{eqnarray*}
t(s_1,w)=\frac{\nu_1+\{s_1-g_1(w)\}^2/\sigma^2_{S_1}}{\nu+1}.
\end{eqnarray*}
Because $\{1,g_1(w), f_1(w),\ldots,f_J(w)\}$ are linearly independent, we can identify $(\beta_0,\ldots,\beta_J,\alpha_1,\alpha_0)$ from the identifiability of the Robit model. Similarly, we can identify $(\beta'_0,\ldots,\beta'_J,\alpha'_1,\alpha'_0)$. Therefore, we can identify the PCEs. \QEDB

\section*{\bf Appendix C: More details for the application}

\subsection*{\bf Appendix C.1: Technical issues with a discrete intermediate variables}

\renewcommand {\theproposition} {S\arabic{proposition}}

\setcounter{table}{0}
\renewcommand {\thetable} {S\arabic{table}}
\setcounter{figure}{0}
\renewcommand {\thefigure} {S\arabic{figure}}

Corollary~\ref{cor:trivariateN} requires $W$ to be continuous but $W$ is categorical in the application. We give the following proposition to formally justify the identifiability of the PCEs in our data analysis.
\begin{proposition}\label{ex:paraY}
Under Assumptions 1 and 2, assume that $\pr(S_1,S_0 \mid W=w)$ is identifiable, and $Y_1$ and $Y_0$ follow linear models
\begin{eqnarray*}
\E(Y_z\mid S_1,S_0,W)=\beta_{z0}+\beta_{z1}S_1+\beta_{z2} S_0, \quad (z=0,1).
\end{eqnarray*}
If there exist $s_0$ and $s_1$ such that neither $\E(S_1\mid S_0=s_0,W=w)$ nor $\E(S_0\mid S_1=s_1,W=w)$ is constant in $w$, then the PCEs are identifiable.
\end{proposition}
\noindent {\it Proof of Proposition~\ref{ex:paraY}.} From the linear model of $Y_0$, 
\begin{eqnarray*}
&&\E(Y\mid Z=0,S=s_0,W=w)\\
&=&\int \E(Y_0\mid S_1=s,S_0=s_0, W=w)\pr(S_1=s \mid S_0=s_0,W=w) \text{d} s\\
&=&\int (\beta_{00}+\beta_{01}S_1+\beta_{02} S_0)\pr(S_1=s\mid S_0=s_0,W=w) \text{d} s\\
&=& \beta_{00}+\beta_{01}\E(S_1\mid S_0=s_0,W=w)+\beta_{02} s_0.
\end{eqnarray*}
Because $\E(S_1\mid S_0=s_0,W=w)$ is not constant in $w$, $\beta_{00}$, $\beta_{01}$ and $\beta_{02}$ are identifiable. Similarly, $\beta_{10}$, $\beta_{11}$ and $\beta_{12}$ are identifiable and hence the PCEs are identifiable. \QEDB

In Proposition \ref{ex:paraY}, the condition that neither $\E(S_1\mid S_0=s_0,W=w)$ nor $\E(S_0\mid S_1=s_1,W=w)$ is constant in $w$ is testable because $\pr(S_1,S_0 \mid W=w)$ is identifiable. In our application, the conditions in Proposition~\ref{ex:paraY} hold and thus the PCEs are identifiable even if the intermediate variable $W$ is categorical.

\subsection*{\bf Appendix C.2: Bayesian analysis with different priors}

\begin{figure}[h]
 \centering
\subfigure{
  \label{fig:jobs1} 
\includegraphics[width=0.65\textwidth]{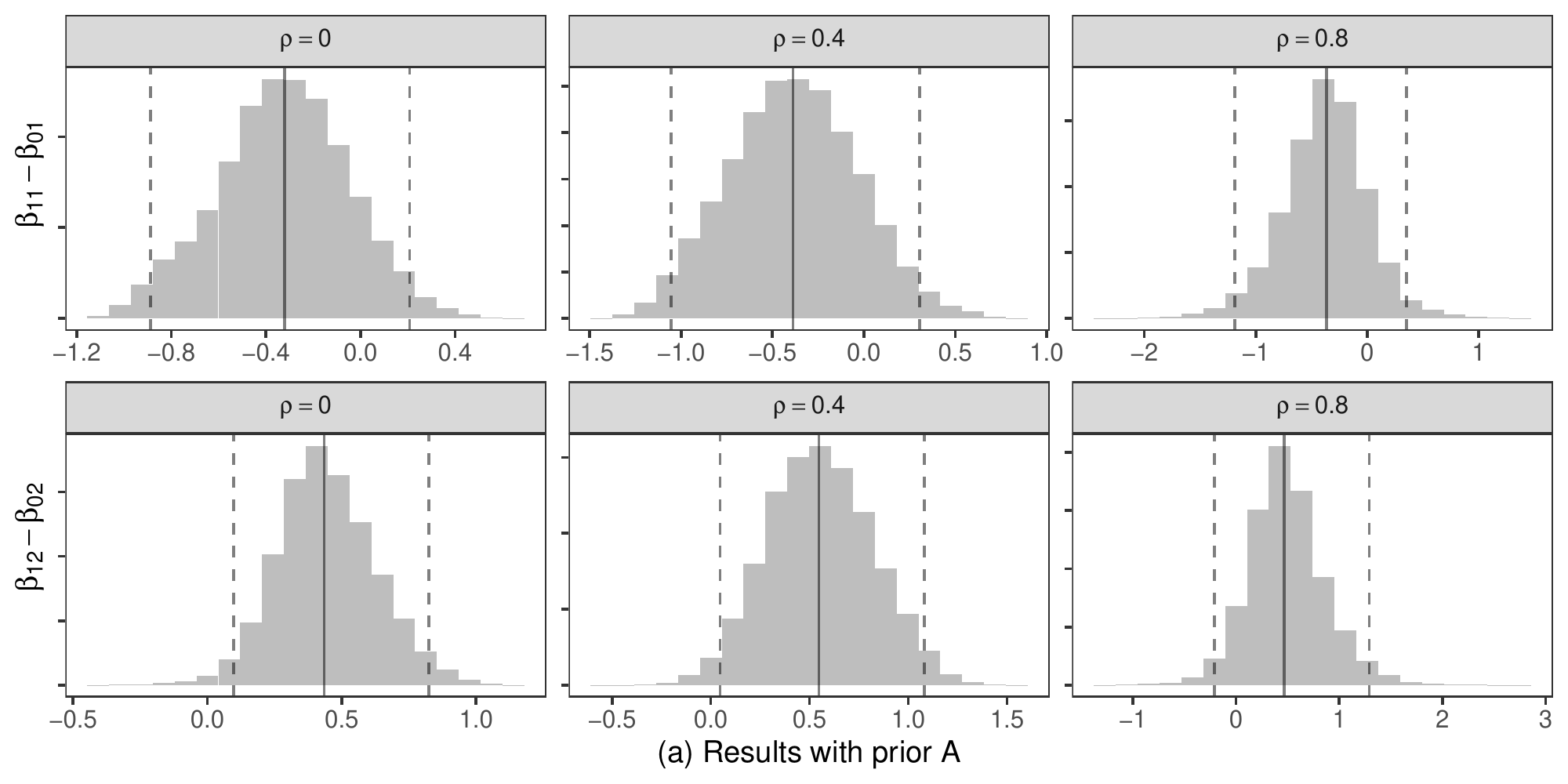}}
\subfigure{
  \label{fig:jobs2} 
\includegraphics[width=0.65\textwidth]{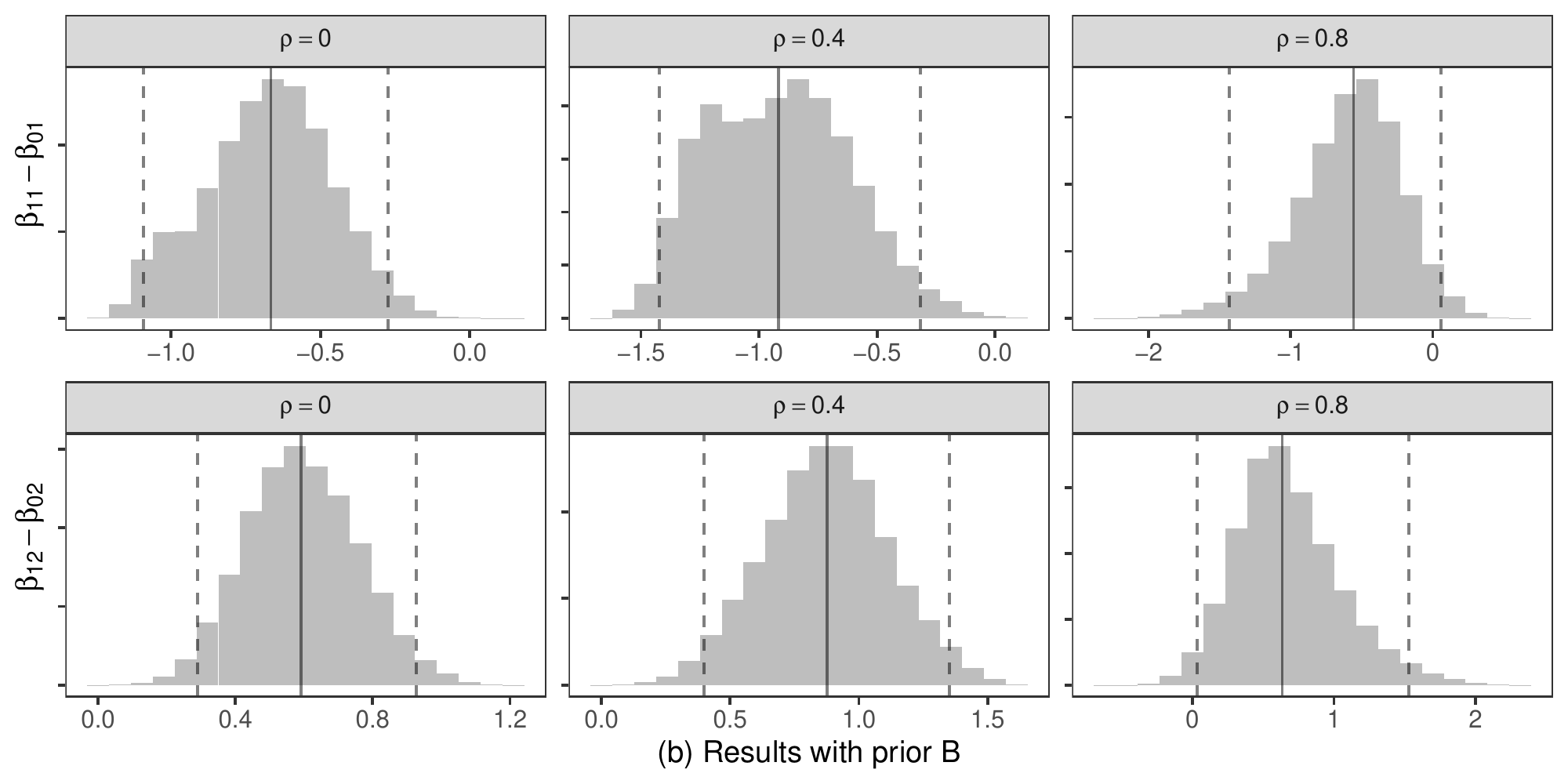}}
 \caption{Histograms of the posterior distributions of $\beta_{11}-\beta_{01}$ and $\beta_{12}-\beta_{02}$. 
The solid lines are the medians and the dashed lines are the posterior 2.5\% and 97.5\% quantiles. 
Prior (A) uses $\bOmega_z=10^2\text{diag}(1,1,1)$ and $\bOmega=10^2\text{diag}(1,1)$, and prior (B) uses $\bOmega_z=\text{diag}(1,1,1)$ and $\bOmega=\text{diag}(1,1)$.
  }
 \label{fig:jobs} 
\end{figure}

We choose two different priors. 
 Let $\bm{\beta}_1=(\beta_{10},\beta_{11},\beta_{12})$, $\bm{\beta}_0=(\beta_{00},\beta_{01},\beta_{02})$ and $\bm{\mu}_{w} =(\mu_1(w),\mu_0(w))$. 
We choose multivariate Normal priors for $\bm{\beta}_z$ and $\bm{\mu}_{w}$: $\bm{\beta}_z \sim \bm{N}_3(\bm{0},\bOmega_z)$, $\bm{\mu}_{w} \sim \bm{N}_2(\bm{0},\bOmega)$ for $w=1,\dots,7$. Prior (A) uses $\bOmega_z=10^2\text{diag}(1,1,1)$ and $\bOmega=10^2\text{diag}(1,1)$ for $z=0,1$; Prior (B) uses $\bOmega_z=\text{diag}(1,1,1)$ and $\bOmega=\text{diag}(1,1)$ for $z=0,1$. We choose the following non-informative parameters for other parameters: $f(\sigma^2_{zw}) \propto 1/ \sigma^2_{zw}$, $f(\sigma^2_{Y_z}) \propto 1/\sigma^2_{Y_z}$, $\{\pr(W=1),\ldots,\pr(W=7)\} \sim \text{Dirichlet}(1,\ldots,1)$ and $\pr(Z=1\mid W=w) \sim \text{Beta}(1,1)$, where $z=0,1$, and $w=1,\dots,7$. Figure \ref{fig:jobs} presents the results which are not sensitive to different priors.

\subsection*{\bf Appendix C.3: More sensitivity analysis for the application }

Section \ref{sec::application} conducts the analysis without covariates. This section includes covariates in the data analysis which can make Assumption~\ref{asm:exclusion} more plausible. Similar to the main text, we will also assess the sensitivity of the results to different values of the correlation coefficient between $S_1$ and $S_0$ given $W$.  

Let $\bm{X}$ be the covariates. We consider the following model for $(S_1,S_0, W)$:
\begin{eqnarray*}
(S_1,S_0)\mid W=w,\bm{X}=\bm{x} \sim \bm{N}_2 \left\{  
\begin{pmatrix}
\mu_1(w,\bm{x})\\
\mu_0(w,\bm{x})
\end{pmatrix},\bm{\Sigma}(w)= \begin{pmatrix}
 \sigma^2_{1}(w)& \rho(w)\sigma_{1}(w)\sigma_{0}(w)\\
\rho(w)\sigma_{1}(w)\sigma_{0}(w) & \sigma^2_0(w)
\end{pmatrix}
 \right\},
\end{eqnarray*}
where $\rho(w)=\rho$, $\mu_1(w,\bm{x})=\alpha_{1w}+\gamma_{1w}^\top\bm{x}$ and $\mu_0(w,\bm{x})=\alpha_{0w}+\gamma_{0w}^\top\bm{x}$. Therefore, we allow the mean of $S_1$ and $S_0$ to depend on the covariates in the sensitivity analysis. We then consider the following model for the potential outcomes: 
\begin{eqnarray*}
 Y_z= \beta_{z1} S_1+\beta_{z2} S_0+ \beta_{z,\bm{X}}\bm{X}+\epsilon_{Y_z}.
\end{eqnarray*}
 
For the simplicity of sensitivity analysis, we use an alternative estimation strategy based on the method of moments, which does not require to specify the distributions of $\epsilon_{Y_1}$ and $\epsilon_{Y_0}$: 
\begin{enumerate}
\item Obtain the estimates of $\bm{\mu}_1(w,\bm{x})$, $\bm{\mu}_0(w,\bm{x})$ and $\bm{\Sigma}(w)$ for all $w$ from the observed distribution of $(S_1,S_0, W,\bm{X})$.
\item For units with $Z_i=1$, impute $S_{i0}$ with $\widehat{S}_{i0}=\widehat{\mu}_{0}(w,\bm{x})+\rho\widehat{\sigma}_{0}(w)/\widehat{\sigma}_{1}(w)\{S_{i1}-\mu_1(w,\bm{x})\}$; for For units with $Z_i=0$, impute $S_{i1}$ with $\widehat{S}_{i1}=\widehat{\mu}_{1}(w,\bm{x})+\rho\widehat{\sigma}_{1}(w)/\widehat{\sigma}_{0}(w)\{S_{i0}-\mu_0(w,\bm{x})\}$.
\item Obtain the estimates of $(\beta_{11},\beta_{12},\beta_{1,\bm{X}})$ by regressing $Y_i$ on $S_{i1}$, $\widehat{S}_{i0}$ and $\bm{X}_i$ for units with $Z_i=1$; obtain the estimates of $(\beta_{01},\beta_{02},\beta_{0,\bm{X}})$ by regressing $Y_i$ on $\widehat{S}_{i1}$, $S_{i0}$ and $\bm{X}_i$ for units with $Z_i=0$.
\end{enumerate}
We use the bootstrap to get the 95\% confidence intervals of the PCEs.  
We first conduct the sensitivity analysis without covariates.
The upper panel of Table~\ref{tab:sens:mom} shows the point estimates and 95\% credible intervals of the five PCEs from the methods of moments. The results are similar to those from the Bayesian approach presented in the main text.

We then consider the sensitivity analysis with covariates. Due to the limited sample size, we only include the demographic variables, gender, age, and race. The lower panel of Table~\ref{tab:sens:mom} shows the point estimates and 95\% credible intervals of the five PCEs from the methods of moments. Although the estimates change in the sensitivity analysis, we can still conclude that for people who can gain more for the job-search self-efficacy from the treatment, the treatment can lower the risk of depression to a larger extent.

\begin{table}[htbp]
\caption{Point and interval estimates of representative PCEs using the methods of moments. The intervals excluding zero are highlighted in bold.}
\begin{center}
\scriptsize
\begin{tabular}{cccccc}
\hline
&   \multicolumn{5}{c}{{\bf Without covariates}} \\
       $(S_1,S_0) $                      &    $\rho$=0  &   $\rho$=0.2   &  $\rho$=0.4   &$\rho$= 0.6   &   $\rho$=0.8   \\
                              \cline{2-6}
$(1.00,5.00)$  &  1.210  & 1.362  & 1.440  &  1.406  &  1.366\\
                      &  $(-0.643,3.093)$  &  $(-1.162,3.717)$ & $(-1.568,4.189)$ & $(-1.328,3.987)$ &$(-0.948,3.463)$\\
 $(3.67,4.50)$  &  $0.243$  &  $0.277$ & $0.398$  &  $0.293$  &  $0.282$\\
                      &  $(-0.082,0.527)$  &  $(-0.202,0.685)$ & $(-0.295,0.817)$ & $(-0.274,0.818)$ &$(-0.246,0.733)$\\
  $(4.17,4.00)$  &  $-0.091$  &  $-0.094$ & $-0.096$  &  $-0.095$  &  $-0.095$\\
                      &  $(-0.189,0.014)$  &  $(-0.200,0.013)$ & $(-0.209,0.013)$ & $(-0.202,0.013)$ &$(-0.196,0.011)$\\
   $(4.67,3.58)$  &  $-0.393$  &  $-0.431$ & $-0.453$  &  $-0.447$  &  $-0.436$\\
                    &  $\bm{(-0.776,-0.005)}$  &  $(-0.916,0.120)$ & $(-1.086,0.211)$ & $(-1.062,0.189)$ &$(-0.961,0.160)$\\
   $(5.00,1.67)$  &  $-1.210$  &  $-1.331$ & $-1.416$  &  $-1.409$  &  $-1.362$\\
                      &  $\bm{(-2.086,-0.317)}$  &  $\bm{(-2.519,-0.105)}$ & $\bm{(-3.096,-0.471)}$ & $(-3.150,0.534)$ &$(-2.975,0.625)$\\
                       \hline \hline
 &   \multicolumn{5}{c}{{\bf With covariates}} \\
       $(S_1,S_0) $                      &    $\rho$=0  &   $\rho$=0.2   &  $\rho$=0.4   &$\rho$= 0.6   &   $\rho$=0.8   \\
                              \cline{2-6}
$(1.00,5.00)$  &  1.581  & 1.704  & 1.732  &  1.466  &  1.562\\
                      &  $\bm{(0.174,2.625)}$  &  $\bm{(0.026, 2.789)}$ & $(-0.030,2.641)$ & $(-0.094,2.531)$ &$(-0.115,2.396)$\\
 $(3.67,4.50)$  &  $0.273$  &  $0.312$ & $0.334$  &  $0.333$  &  $0.317$\\
                      &  $\bm{(0.026,0.454)}$  &  $\bm{(0.004,0.493)}$ & $(-0.029,0.520)$ & $(-0.039,0.521)$ &$(-0.057,0.518)$\\
  $(4.17,4.00)$  &  $-0.103$  &  $-0.105$ & $-0.104$  &  $-0.101$  &  $-0.098$\\
                      &  $(-0.202,0.008)$  &  $(-0.204,0.007)$ & $(-0.200,0.005)$ & $(-0.200,0.006)$ &$(-0.195,0.008)$\\
   $(4.67,3.58)$  &  $-0.452$  &  $-0.490$ & $-0.506$  &  $-0.498$  &  $-0.476$\\
                    &  $\bm{(-0.690,-0.131)}$  &  $(-0.744,-0.101)$ & $\bm{(-0.757,-0.083)}$ & $\bm{(-0.749, -0.061)}$ &$\bm{(-0.772,-0.040)}$\\
   $(5.00,1.67)$  &  $-1.213$  &  $-1.376$ & $-1.485$  &  $-1.506$  &  $-1.462$\\
                      &  $\bm{(-1.788,-0.389)}$  &  $\bm{(-2.002,-0.306)}$ & $\bm{(-2.093,-0.225)}$ & $\bm{(-2.182,-0.142)}$ &$\bm{(-2.169,-0.089)}$\\
                       \hline

\label{tab:sens:mom}

\end{tabular}
\end{center}
\end{table}

\end{document}